%% file: main.tex
\documentclass[longauth]{aa}

\usepackage{graphicx}

\usepackage{booktabs}
\usepackage{multicol}
\usepackage{multirow}

\usepackage{float}
\usepackage[section]{placeins} % Used for \Floatbarriers command used in the appendix

\usepackage{txfonts}
\usepackage{hyperref}

\usepackage{natbib}
\usepackage{siunitx}
\usepackage{csquotes}

\usepackage{txfonts} % Provides \star, might want to replace

\usepackage[version=4]{mhchem}

\newcommand{\zenododoi}{\href{https://doi.org/10.5281/zenodo.19356398}{10.5281/zenodo.19356398}}
\newcommand{\comment}[2][{[}...{]}]{\textcolor{red}{#1}}

\makeatletter
\newcommand{\showfontsize}[1]{%
  \begingroup
  #1%
  \typeout{Font size for \string#1 is \f@size pt (baseline skip: \f@baselineskip pt)}%
  \endgroup
}
\makeatother

\begin{document}

\title{Information content of JWST transmission spectroscopy of the exoplanet HAT-P-12b from the optical to the mid-infrared} % TODO: Use custom commands here as well?

\titlerunning{Information Content of JWST Transmission Spectroscopy of HAT-P-12b}
\authorrunning{L. Heinke et al.}

\input{authors}
\input{affiliations}
\date{Received ...; accepted ...}

% Use of structured abstract not mandatory, also possible to follow this implicitly
% Only "Aims", "Methods", and "Results", are mandatory when following structured approach
% Never more than 1-3 sentences for first sections, results and conclusions maybe longer, perhaps also methods
\abstract{
    % Context
    % Can also contain a bit of motivation such as why was this done/why relevant
    The James Webb Space Telescope (JWST) provides low- to medium-resolution spectra with unprecedented precision and broad near- to mid-infrared wavelength coverage, enabling the detailed characterization of exoplanet atmospheres. Given the complexity of JWST data and the diversity of observing modes, it is essential to understand the information content of the resulting spectra to optimize observation strategies and assess the limits of atmospheric inference.
}{
    % Aims
    % Can also contain some basic information about the target
    We present a new JWST NIRISS SOSS transit observation of the warm sub-Saturn \object{HAT-P-12b}. Together with complementary NIRSpec G395M and MIRI LRS data, this enables a detailed assessment of the information content across JWST instruments over the full accessible wavelength range (excluding MIRI MRS).
}{
    % Methods
    % Should be on both
    The NIRISS data were reduced, and the impact of specific reduction choices on the resulting transmission spectrum was assessed. Atmospheric retrievals were performed for all combinations of JWST data, supplemented by archival HST observations in select cases. The analysis further included evaluations of molecular detection significances and assumptions about the atmospheric structure.
}{
    % Results
    % More about basic findings not so much implications
    The same four molecules previously reported were significantly detected: \ce{H2O}, \ce{CO2}, \ce{CO}, and \ce{H2S}.
    Except for \ce{H2O}, all required NIRSpec coverage for detection, while \ce{H2S} was only detected in multi-instrument retrievals. Abundance constraints obtained using HST WFC3 instead of JWST NIRISS SOSS were largely consistent, particularly when combining instruments, but NIRISS SOSS proved essential to establish robust evidence for non-gray cloud behavior. A scattering slope of moderate steepness ($p < 4$) was consistently retrieved, independent of the chosen HST STIS reduction.
}{
    % Conclusions
    % Overview of what was learned form this study and what broader implications are
    Single-instrument retrievals, even when yielding significant detections, tend to overestimate molecular abundances. In contrast, retrievals that combine spectra from multiple JWST instruments generally converge toward consistent abundance constraints. The derived C/O ratio remains sensitive to subtle differences between NIRSpec reductions, owing to the instrument’s exclusive coverage of the carbon-bearing molecules \ce{CO2} and \ce{CO}, so its interpretation requires caution. The results are broadly consistent with information content studies of the benchmark target WASP-39b, although differences, including the absence of a preference for non-isothermal $T$--$P$ profiles, highlight variations in information content across exoplanet types.
}

\maketitle

\showfontsize{\tiny}
\showfontsize{\scriptsize}
\showfontsize{\footnotesize}
\showfontsize{\small}
\showfontsize{\normalsize}
\showfontsize{\large}
\showfontsize{\Large}
\showfontsize{\LARGE}
\showfontsize{\huge}
\showfontsize{\Huge}

\section{Introduction}\label{sec:introduction}

Advances in instrumentation have enabled exoplanet science to progress from large-scale detection to the early stages of large scale atmospheric characterization. This is exemplified by the over 300 planets\footnote{\textit{Exoplanet Atmospheres Database}: \url{https://research.iac.es/proyecto/exoatmospheres/index.php}; accessed on 2 March 2026} whose atmospheres have been characterized to some extent. Because most planets cannot be observed directly, their spectral signatures must be disentangled from those of their host stars. Arguably the most important technique for doing this is transmission spectroscopy \citep{Deming_2019_HowCharacterizeAtmosphere}, which measures the wavelength dependence of the transit depth that arises in the presence of an atmosphere.
Ground-based low-resolution transmission spectroscopy is feasible in the optical \citep[e.g.][]{Nikolov_2016_VLTFORS2Comparative} and, to a limited extent, has even been applied up to the long wavelength end of the near-infrared \citep[\SI{\sim 2.4}{\micro\meter};][]{Crossfield_2013_WarmIceGiant, Danielski_2014_094242MmGroundbased}.
Performing ground-based astronomical measurements further into the infrared (IR) is prevented by the obstructing influence of the Earth's atmosphere. Probing these wavelengths, however, is crucial for characterizing the chemical composition of atmospheres, as most molecular ro-vibrational transitions occur in this region.
Performing transmission spectroscopy from space can mitigate this problem. Before JWST, the primary instrument used for near-IR transmission spectroscopy was the Hubble Space Telescope's (HST) WFC3, mainly utilizing its G141 grism. Its \SIrange{1.1}{1.7}{\micro\meter} wavelength range enabled the detection of water from two prominent absorption bands \citep[e.g.][]{Iyer_2016_CharacteristicTransmissionSpectrum, Tsiaras_2018_PopulationStudyGaseous} as well as the feature-muting effect of clouds \citep{Sing_2016_ContinuumClearCloudy, Estrela_2022_TemperatureTrendClouds}. Additionally, the data were useful for establishing upper limits on methane abundances \citep{Benneke_2019_SubNeptuneExoplanetLowmetallicity, Carone_2021_IndicationsVeryHigh, Edwards_2023_ExploringAbilityHubble}, although the instrument mode's narrow wavelength range and the overlapping opacity bands of water and methane led to degeneracies in the retrieved abundances of both molecules \citep{Bezard_2022_MethaneDominantAbsorber}. The direct detection of carbon-bearing molecules, however, remained challenging \citep{Changeat_2020_KELT11AbundancesWater, Edwards_2023_CharacterizingWorldHotNeptune}. Measurements at longer wavelengths were only possible with the Spitzer Space Telescope and were mostly performed using the two photometric channels that remained operational after the end of the cryogenic mission, the IRAC instrument's \SIlist{3.6; 4.5}{\micro\meter} bands \citep[for review see][]{Deming_2020_HighlightsExoplanetaryScience}. These were only able to provide tentative hints of possible features of molecules such as \ce{CO2} and \ce{CO} \citep{Benneke_2019_SubNeptuneExoplanetLowmetallicity, Spake_2021_AbundanceMeasurementsH2O, Alderson_2022_ComprehensiveAnalysisWASP17bs}.

The advent of the James Webb Space Telescope (JWST) marked a significant advance in observational capabilities in the infrared. Using a combination of instruments, spectra can be obtained with a continuous wavelength coverage from \SI{\sim 0.6}{\micro\meter} up to \SI{\sim 12}{\micro\meter} with MIRI LRS, and extending to \SI{\sim 28}{\micro\meter} with MIRI MRS \citep{Gardner_2023_JamesWebbSpace}. The use of MIRI MRS for transmission spectroscopy has been limited to date, with only a single exoplanet observed during the first four JWST cycles \citep{Deming_2024_ExoplanetTransitSpectroscopy}.
However, most of the other suitable JWST modes have been applied extensively for transmission spectroscopy, leading to several notable discoveries \citep[see][for an overview]{Espinoza_2025_HighlightsExoplanetObservations}. A number of molecules, previously only detected at low significance, have now been confirmed with greater certainty, while others have been detected for the first time. These include dominant carbon species such as \ce{CO2} \citep{JWSTTransitingExoplanetCommunityEarlyReleaseScienceTeam_2023_IdentificationCarbonDioxide, Mayo_2025_DetectionH2OCO2}, \ce{CO} \citep{Rustamkulov_2023_EarlyReleaseScience, Sing_2024_WarmNeptunesMethane}, and \ce{CH4} \citep{Madhusudhan_2023_CarbonbearingMoleculesPossible, Bell_2023_MethaneThroughoutAtmosphere}, the important sulfur reservoirs \ce{H2S} \citep{Thao_2024_FeatherweightGiantUnraveling} and \ce{SO2} \citep{Alderson_2023_EarlyReleaseScience, Gressier_2024_JWSTs065um}, and less commonly detected molecules such as the nitrogen-bearing molecule ammonia (\ce{NH3}) \citep{Welbanks_2024_HighInternalHeat} and even \ce{SiO} \citep{Ma_2025_NewLookAtmospheric, Gapp_2025_WASP121BsTransmission}. The presence of \ce{SO2} in detectable quantities has been interpreted as the first direct evidence of photochemistry \citep{Tsai_2023_PhotochemicallyproducedSO$_2$Atmosphere}. The broader wavelength coverage of JWST enables tighter constraints on cloud-induced spectral slopes, revealing evidence of complex, non-gray cloud behavior \citep{Feinstein_2023_EarlyReleaseScience, Lueber_2024_InformationContentJWST, Roy-Perez_2025_RoleCloudParticle}. In a few cases, features in the mid-IR have even been directly attributed to cloud particles composed of silicates \citep{Grant_2023_JWSTTSTDREAMSQuartz, Dyrek_2024_SO2SilicateClouds}.

\object{HAT-P-12b} \citep{Hartman_2009_HATP12bLowDensitySubSaturn} is a warm sub-Saturn \citep[$T_\mathrm{eq} \sim \SI{960}{\kelvin}$, $R_\mathrm{p} \sim 0.95 \, \mathrm{R_{Jup}}$, $M_\mathrm{p} \sim 0.21 \, \mathrm{M_{Jup}}$][and hereafter]{Akeson_2013_NASAExoplanetArchive} that has been the focus of numerous studies since its discovery. Focusing on transmission spectroscopy, the combination of the planet's low density ($\rho \sim \SI{0.3}{\gram\per\cubic\centi\meter}$), the large planet-to-star radius ratio ($R_\mathrm{p} / R_\star \sim 0.14$) and its bright host star (K4/K5 V; $m_\mathrm{J} \sim 10.8$) makes HAT-P-12b a highly suitable target for this method, as reflected by its high transmission spectroscopy metric  \citep[TSM; defined by][]{Kempton_2018_FrameworkPrioritizingTESS} of $\sim 325$ \citep{Nikolov_2022_TrExoLiSTSTransitingExoplanets}.
This led to the detection of \ce{H2O} via the HST WFC3 observations, tentative hints of a possible \ce{CO2} feature from the Spitzer photometry, and the general feature-muting imprint of clouds \citep{Wong_2020_OpticalNearinfraredTransmission}.
The HST STIS data were initially interpreted as showing a strong Rayleigh scattering slope \citep{Sing_2016_ContinuumClearCloudy}, which was incompatible with previous ground-based observations indicating a flat optical spectrum \citep{Mallonn_2015_BroadbandSpectrophotometryHot}. \citet{Alexoudi_2018_DecipheringAtmosphereHATP12b} demonstrated that a moderate steepness sub-Rayleigh slope could be inferred for both datasets when performing a consistent reduction. Subsequently obtained ground-based photometry supported this interpretation \citep{Yan_2020_LBTTransmissionSpectroscopy}. The behavior in the optical has, however, also been interpreted as showing potential signs of stellar contamination \citep{Jiang_2021_EvidenceStellarContamination}.
\citet{Crouzet_2025_DetectionCO$_2$CO} presented the analysis of a JWST NIRSpec G395M transmission spectrum which, when combined with shorter-wavelength HST WFC3 data, led to significant detections of \ce{H2O}, \ce{CO2}, and \ce{CO}, a non-detection of \ce{CH4}, as well as a tentative detection of \ce{H2S}. 

This study introduces the JWST NIRISS SOSS transmission spectrum of HAT-P-12b. This extends the spectral coverage to the shortest wavelengths accessible with JWST (\SI{\sim 0.6}{\micro\meter}), beyond the range of HST WFC3 and down to the upper wavelength limit of HST STIS, probing the infrared-to-optical transition where the onset of the elusive optical slope is expected.
Taken together with additional observations with NIRSpec BOTS G395M \citet{Crouzet_2025_DetectionCO$_2$CO} and MIRI LRS (Bouwman et al., in prep.), the resulting dataset is part of a smaller subset of planets for which JWST transmission spectroscopy has been taken over the entire feasible wavelength range \citep[17 planets up to cycle 4, corresponding to about $\SI{12}{\percent}$ of targets;][]{Nikolov_2022_TrExoLiSTSTransitingExoplanets}\footnote{\textit{TrExoLiSTS}: \url{https://www.stsci.edu/~nnikolov/TrExoLiSTS/JWST/trexolists.html}}, not counting MIRI MRS, which has thus far only seen very limited use in the characterization of exoplanet atmospheres \citep{Deming_2024_ExoplanetTransitSpectroscopy}.
JWST observations of the planet \object{WASP-39b} \citep{Carter_2024_BenchmarkJWSTNearinfrared} have served as a benchmark dataset for assessing the information content of JWST transmission spectra \citep[e.g.][]{Lueber_2024_InformationContentJWST, Fisher_2024_JWSTNIRISSHST}. By evaluating various instrument combinations and testing the capabilities of  JWST for a different class of exoplanet, this study contributes to a broader understanding of its capabilities across the full accessible wavelength range and multiple planetary types. The paper is structured as follows: Section~\ref{sec:observations} provides an overview of the observations used in this study. Section~\ref{sec:data_reduction} explains how the data reduction of the NIRISS SOSS dataset was conducted. Section~\ref{sec:retrievals} describes the used retrieval setup. Results are presented in Section~\ref{sec:results}, followed by a discussion in Section~\ref{sec:discussion}, and conclusions in Section~\ref{sec:conclusions}.

\section{Observations}\label{sec:observations}
An overview of the observations analyzed in this study is provided in Table~\ref{tab:observations}.
All three JWST transit spectra were obtained as part of a Guaranteed Time Observation (GTO) program of the ExoMIRI subgroup of the MIRI European Consortium \citep{Lagage_2017_MIRINIRSPECTransit}. The NIRISS data were obtained in the NISRAPID readout pattern applied to the SUBSTRIP256 subarray using 15 groups per integration. The total observation time was $\sim 6.2$ hours which corresponds to 254 integrations. This covered the $\sim 2.3$ hour transit along with a stellar baseline of $\sim 2.4$ hours before ingress and $\sim 1.4$ hours after egress. No additional exposure using the F277W filter was obtained. By restricting the aperture position angle,
%to \SIrange{292}{180}{\degree}
any major contaminations could be avoided, rendering such an exposure non-critical. Details on the NIRSpec and MIRI data and their reductions are provided in \citet{Crouzet_2025_DetectionCO$_2$CO} and Bouwman et~al. (in prep.). This study primarily uses the reductions obtained with CASCADe, while the TEATRO reduction of the NIRSpec data is employed for additional robustness tests.

\begin{table*}\label{tab:observations}
\caption{Key properties of the observations and resulting transmission spectra.}
\begin{center}
\input{tables/observations}
\end{center}
\tablefoot{PID = Program ID. The ranges of the transit depth uncertainties $\sigma_\delta$ correspond to the central \SI{95}{\percent} credible intervals. Data reductions: [0] This work, [1] \citet{Crouzet_2025_DetectionCO$_2$CO}, [2] Bouwman et~al. (in prep.), [3] \citet{Sing_2016_ContinuumClearCloudy}, [4] \citet{Alexoudi_2018_DecipheringAtmosphereHATP12b}.}
\end{table*}

In select cases, HST data were also included for further analysis. The spatial scanning mode HST WFC3 data first presented by \citet{Tsiaras_2018_PopulationStudyGaseous} was used.
%, which is not to be confused with the lower-quality earlier scanning mode observations \citep{Line_2013_NearInfraredTransmissionSpectrum, Sing_2016_ContinuumClearCloudy}. 
As in the other papers of this series, this study uses the CASCADe reanalysis of the data (see \citealt{Crouzet_2025_DetectionCO$_2$CO} for details). In addition, HST STIS data were used to provide information further into the optical. No reanalysis of the data were performed. Instead, the impact of the two distinct reductions presented in \citet{Sing_2016_ContinuumClearCloudy} and \citet{Alexoudi_2018_DecipheringAtmosphereHATP12b} were tested.

\section{NIRISS data reduction}\label{sec:data_reduction}
The reduction of transmission spectroscopy data requires transforming raw detector readouts into calibrated spectra and fitting the resulting transit lightcurves to derive the transit spectrum. Several methodological choices in this process can significantly affect the inferred atmospheric properties \citep{Mugnai_2024_ComparingTransitSpectroscopy}.
As the true planetary signal cannot be directly observed, it is often unclear which reduction approach most faithfully recovers it.
Even seemingly minor differences in the resulting transmission spectrum can impact the inferred atmospheric composition \citep[e.g.][]{Crouzet_2025_DetectionCO$_2$CO}.
One commonly adopted approach to test the robustness of the obtained transmission spectra is to perform multiple independent data reductions using different pipelines \citep[e.g.][]{Feinstein_2023_EarlyReleaseScience, Rustamkulov_2023_EarlyReleaseScience, Radica_2023_AwesomeSOSSTransmission}. This paper presents an analysis similar to that of \citet{Holmberg_2023_ExoplanetSpectroscopyJWST}, 
focusing on the effect of individual \enquote{tweaks} to the data-reduction.
Rather than comparing sets of multiple tweaks, as is effectively done when comparing outcomes of different pipelines, the emphasis here is on isolating the impact of specific steps. The following two sections describe the choices that define the standard reduction, i.e. the reduction used in the subsequent retrieval analysis.
During the first part of the data reduction, the lightcurve extraction, the choices made mostly followed the default methods of the used pipeline.
In the second part of the reduction, an approach similar to that applied to the other two datasets (\citealt{Crouzet_2025_DetectionCO$_2$CO}; Bouwman et~al., in prep.) was adopted, and simplifying assumptions were made whenever applicable.
The applied tweaks and their impacts are described and discussed in Appendix~\ref{app_sec:reduction_tweaks}.

\subsection{Lightcurve extraction}\label{ssec:lightcurve_extraction}

The ExoTEDRF pipeline\footnote{\url{https://github.com/radicamc/ExoTEDRF}} (formerly known as supreme-SPOON; \citealt{Radica_2024_ExoTEDRFEXOplanetTransit}) was used to obtain spectral lightcurves from the uncalibrated group-level images. The pipeline was initially developed to deal with the unique peculiarities of the NIRISS SOSS mode. These include the presence of multiple spectral orders and the non-constant background with its distinct \enquote{step} caused by dispersed zodiacal light. It has been used for NIRISS SOSS transit spectroscopy observations of multiple targets and has also been benchmarked against other pipelines \citep[e.g.][]{Radica_2023_AwesomeSOSSTransmission, Feinstein_2023_EarlyReleaseScience}. 
%The pipeline has more recently also been extended to work with NIRSpec BOTS, with future support for MIRI LRS planned as well \citep{Radica_2024_ExoTEDRFEXOplanetTransit}.

The data reduction starts out from uncalibrated data produced by version \texttt{2023\_1a} of the JWST Science Data Processing (SDP) system. These are then processed using version \texttt{2.0.0} of the ExoTEDRF pipeline. It leverages the standard \texttt{jwst} pipeline for many of the common reduction steps. Version \texttt{1.12.5} \citep{bushouse_2023_10022973} of the \texttt{jwst} pipeline was used, which is the version recommended for the used version of ExoTEDRF. No further refinements were applied to the reference files, which were obtained with the Calibration Reference Data System (CRDS) \citep{Greenfield_2016_CalibrationReferenceData} using context \texttt{1290}. The context version was fixed to ensure a consistent comparison between the different reduction tweaks. Some of the custom ExoTEDRF steps do require additional observation-specific reference files. While these can be estimated dynamically during pipeline execution, more accurate results are typically obtained by first running the full pipeline in a first pass and generating the reference files from the output. This two-pass approach was therefore adopted here.
The setup adopted for the standard reduction closely follows that of \citet{Radica_2023_AwesomeSOSSTransmission}, with one key difference: The \texttt{OneOverF} step is applied not only at the group level but also at the subsequent integration level. Applying the correction at the integration level alone is a commonly adopted approach in NIRISS reductions with other pipelines \citep[e.g.][]{Feinstein_2023_EarlyReleaseScience}. The correction is applied at both stages in the standard reduction to account for the possibility of residual 1/f noise persisting at the integration level after group-level correction, and to define a consistent baseline against which tweaked reductions, where the correction is applied only at a single stage, can be directly compared.
Technical details regarding the pipeline's implementation can be found in the official documentation\footnote{\url{https://exotedrf.readthedocs.io}}.
The ExoTEDRF pipeline can be configured via a \texttt{YAML} configuration file. The file used in the standard reduction is publicly available (see Acknowledgements).

During the first two stages, calibrations and corrections are applied to the data, first at the group level and then, after fitting the integration \enquote{ramps}, to the resulting integration-level images.
%The only tweaks applied here concern the treatment of background removal.
%In this context, \enquote{background} refers to unwanted flux present in the science image that does not originate from the astrophysical target.
The NIRISS background consists of at least two distinct components: the true sky background, primarily composed of dispersed zodiacal light, which enters the detector due to the slitless design of the spectrograph, and the so-called \enquote{$1/f$ noise}. The latter is a form of correlated detector readout noise that affects all JWST NIR detectors to some extent \citep{Schlawin_2020_JWSTNoiseFloor}. Since the $1/f$ noise is much weaker than the sky background, the latter must be removed first. In principle, the $1/f$ noise also varies from group to group and should therefore be removed during the first stage. The sky background on the other hand is best removed later in the second stage close to the spectral extraction. In the reduction approach presented in \citet{Radica_2023_AwesomeSOSSTransmission} this is done by temporarily removing an estimate of the sky background in the first stage to reveal the remaining $1/f$ noise. This noise is then estimated on a column-by-column basis and removed. The sky background gets re-added and then again estimated and subtracted, this time permanently, in the second stage. The present work modifies this approach for the standard reduction by introducing an additional subtraction of a column-specific noise signal in the second stage.
To estimate the sky background, a background model was scaled to the flux level of a median stack of all out-of-transit images, computed either per group or as a single stack at the integration level.
The used model background was obtained during commissioning\footnote{\url{https://stsci.app.box.com/s/9qohomufvzf84tn4gjawnko11oruq2jw}} (formerly listed on JDox; observation 5 of program ID 1541).
The default ExoTEDRF method applied in the standard reduction determines the scaling factor using a single rectangular low-flux region. The default positioning of this region was confirmed to be adequate ($x_1 \in [350, 549]$, $y_1 \in [230, 249]$; see Fig.~\ref{fig:backs+contams}).
To estimate the $1/f$ noise, ExoTEDRF provides four built-in methods. In the standard reduction, the default and simplest one, \texttt{scale-achromatic}, was used. It scales the median-stack images to the estimated flux level of each group or integration image. As the name suggests, the scaling is performed achromatically using a white lightcurve estimate derived from the first pass reduction.

In the third stage, a single step gets executed which extracts a timeseries of spectra from the corrected and calibrated integration images.
Flux extraction was performed with a simple box extraction using the default aperture width of 35 pixels, close to the \enquote{optimal} value of 36 determined by the pipeline’s optimization routine, which selects the aperture width that minimizes white lightcurve scatter.
For determining the trace positions, the optional reference file from the first pass reduction was used. The ExoTEDRF implementation of the spectral extraction step includes an optional wavelength calibration based on theoretical PHOENIX stellar atmosphere models \citep{Husser_2013_NewExtensiveLibrary}. This calibration was applied using stellar parameters ($T_\mathrm{eff}^{\star}$, $\log g_\star$, $[\mathrm{Fe/H}]$) taken from \citet{Mancini_2018_GAPSProgrammeHARPSN} (see also Table~\ref{tab:reduction_params}).

\subsection{Lightcurve fitting}\label{ssec:lcurve_fitting}
Although the ExoTEDRF pipeline \citep{Radica_2024_ExoTEDRFEXOplanetTransit} provides built-in tooling for performing spectral lightcurve fits on its stage 3 output, a custom script based on the same underlying exoplanet system modeling tool, \texttt{juliet} \citep{Espinoza_2019_JulietVersatileModelling}, was used instead.
%More recently, the used methodology has been moved to a stand-alone dedicated framework for performing the lightcurve fitting, exoUPRF \citep{Radica_2024_RadicamcExoUPRFV101}.
This was done to allow for greater flexibility in the chosen fitting approach.
The scripts used to perform the fitting are publicly available (see Acknowledgements).
Overall, an approach similar to that utilized for the TEATRO reductions of the other two HAT-P-12b JWST datasets 
(\citealt{Crouzet_2025_DetectionCO$_2$CO}; Bouwman et~al., in prep.)
was adopted in the standard reduction.

Following well-established methodology \citep[e.g.][]{Stevenson_2014_TransmissionSpectroscopyHot, Kreidberg_2015_DetectionWaterTransmission}, the data are first fitted achromatically using so-called white lightcurves. Two white lightcurves are constructed. The first is derived from the full wavelength range of the first spectral order (\SIrange{0.85}{2.83}{\micro\meter}).
The second white lightcurve is derived from the second spectral order, which nominally spans approximately \SIrange{0.59}{1.41}{\micro\meter}. However, the data at both ends of this range are too noisy and are therefore excluded from further analysis. In addition, the longer-wavelength part of the second order overlaps with the first order, which provides significantly higher throughput (up to a factor of $\sim 500$ at \SI{1.27}{\micro\meter}). For these reasons, only a restricted wavelength range of \SIrange{0.6}{0.85}{\micro\meter} is used for the second order white lightcurve.
The resulting lightcurves are normalized by the mean out-of-transit baseline flux. The first integration was excluded from the baseline normalization and subsequent fitting, as it is clearly identifiable as an outlier (see Fig.~\ref{fig:white_lcurves}); no binning in the temporal domain was applied.
Both lightcurves are fitted simultaneously using a single combined model that shares the same orbital parameters.
Either wide uninformative priors or Gaussian priors based on previous literature findings were used. The full set of used parameters, including priors and posterior values are listed in Table~\ref{tab:reduction_params}.
There are several possible parametrizations of the orbit available \citep{Espinoza_2019_JulietVersatileModelling}. Here, a parametrization based on the orbital period $P$, transit mid-time $t_0$, scaled semi-major axis $a/R_\star$, impact parameter $b$, eccentricity $e$, and argument of periastron $\omega$ was adopted.
The orbital period was fixed to the literature value reported by \citet{Kokori_2022_ExoClockProjectII}. We assumed a circular orbit, fixing the eccentricity to $e=0$ (consistent with the $1\sigma$ upper limit $e<0.035$ reported by \citealt{Bonomo_2017_GAPSProgrammeHARPSN}). In this case, the argument of periastron is undefined and was therefore arbitrarily fixed to $\omega=\SI{90}{\degree}$.
The two white lightcurves are constructed by binning the timeseries of spectra over distinct wavelength ranges, resulting in variations in transit shape due to the wavelength dependence of both the transit depth and limb darkening. Both effects therefore need to be treated separately for each spectral order. For the former, this implies fitting two independent transit depths $\delta$.
Limb darkening is likewise freely fitted for to account for known degeneracies with orbital parameters that also influence the transit shape, such as the impact parameter $b$ \citep{Espinoza_2015_LimbDarkeningExoplanets}. Specifically, a quadratic law was adopted. The limb darkening coefficients (LDCs) $u_1$ and $u_2$ were permitted to vary freely (hereafter referred to as \enquote{free}) within the physically informed reparametrization proposed by \citet{Kipping_2013_EfficientUninformativeSampling}, expressed in terms of the transformed parameters $q_1$ and $q_2$.
The baseline was assumed to be flat.
The \texttt{juliet} transit lightcurve model also includes several order-specific systematic parameters, namely the dilution factor $D$, the relative out-of-transit flux $M$, and a jitter term $\sigma_\omega$.
The dilution factor was fixed at 1 (no contamination), while the other two parameters were allowed to vary freely within wide uninformative priors.
Together with these additional systematic parameters, a total of 13 parameters were fitted.
The \texttt{juliet} code supports several different samplers. In this work, the \texttt{dynesty} sampler \citep{Speagle_2020_DYNESTYDynamicNested}, a commonly adopted choice \citep[e.g.][]{Pontoppidan_2022_JWSTEarlyReleasea, Fournier-Tondreau_2024_NearInfraredTransmissionSpectroscopy}, was employed with the number of live points set to 5000.

To obtain the final product of the data reduction, the transmission spectrum, the spectral lightcurves first need to be fitted. Following the approach of \citet{Radica_2023_AwesomeSOSSTransmission}, the data were re-binned to a spectral resolution of $R \sim 125$. The bins correspond to full pixel ranges without inter-pixel interpolation (order 1: 4 -- 23 pixels, order 2: 6 -- 15 pixels), resulting in slight variations in spectral resolution between bins (see Table~\ref{tab:observations}). The second spectral order was restricted to the same range as for the white lightcurve.
As in the case of the white lightcurves, the first integration is discarded and no temporal binning is applied.
The orbital parameters were fixed to the values from the white lightcurve fit (see Table~\ref{tab:reduction_params}).
Since the aforementioned degeneracy between these parameters and the LDCs is therefore resolved, and given the much reduced S/N of the spectral lightcurves when compared to the white lightcurves, the LDCs were fixed to values predicted by stellar atmosphere models (hereafter referred to as \enquote{fixed}) instead of freely fitted for. For this purpose, the \texttt{ExoTiC-LD} Python package \citep{Grant_2024_ExoTiCLDThirtySeconds}, together with the Stagger model grid \citep{Magic_2015_StaggergridGrid3D}, was used with the order-specific throughput data from the CRDS NIRISS reference file, evaluated over the wavelength range of each spectral bin.
This leaves a total of three parameters ($R_\mathrm{p}/R_\star$, $M$, $\sigma_\omega$) to be fitted separately for each spectral bin. Reflecting the reduced model complexity and lower S/N of the spectral bins, the number of live points was reduced to 500. Finally, the transmission spectrum is obtained by transforming the marginal posterior of $R_\mathrm{p}/R_\star$ into a posterior of the transit depth $\delta$, adopting the posterior median as the nominal value and the standard deviation as the uncertainty. 
The resulting transmission spectrum is publicly available (see Acknowledgements).

\section{Retrievals}\label{sec:retrievals}

The transmission spectra were fitted using the forward modeling and retrieval framework ARCiS\footnote{\url{https://github.com/michielmin/ARCiS}}. It offers a high degree of flexibility, supporting a wide range of modeling complexities \citep{Min_2020_ARCiSFrameworkExoplanet}.
%These range from simple parametric treatments of the temperature–pressure ($T$--$P$) profile, chemistry, and clouds to self-consistently computed radiative-convective equilibrium, disequilibrium chemistry by vertical mixing \citep{Kawashima_2021_ImplementationDisequilibriumChemistry}, and a physically motivated cloud formation scheme \citep{Ormel_2019_ARCiSFrameworkExoplanet, Huang_2024_ExoLynGoldenMean}.
The code has been extensively tested and benchmarked against other atmospheric retrieval frameworks \citep{Barstow_2022_RetrievalChallengeExercise}.
The general setup closely follows that of the two companion studies (\citealt{Crouzet_2025_DetectionCO$_2$CO}; Bouwman et~al., in prep.). Values from \citet{Mancini_2018_GAPSProgrammeHARPSN} were adopted for the system parameters ($R_\star$, $T_\mathrm{eff}^\star$, $d$, $a$, $M_\mathrm{p}$; see Table~\ref{tab:reduction_params}). The reference pressure $P_\mathrm{ref}$, the pressure level that corresponds to the retrieved planetary radius, was fixed at \SI{10}{\bar}. The atmospheric structure was modeled using 100 layers equidistant in log-space, spanning the range of $[10^{2},10^{-10}]$ \si{\bar}.
%As was shown in \citep{Lueber_2024_InformationContentJWST} the complexity of JWST data, can justify the use of more complex temperature structures beyond a simple isothermal profile. This will even more so be the case when combining datasets from multiple JWST instruments.
The atmospheric structure reflects the complexity of JWST data through an $N$-point profile of five temperature points $T_{\mathrm{p},i}$, evenly spaced in log-pressure over the full modeled pressure range. A smoothing window, averaging over about nine adjacent pressure points ($|\log(P_i/P_j)| \leq 0.6$), yields the final $T$–$P$ profile.
Given the planet's low density, a primordial atmosphere was assumed, i.e. one primarily composed of a background gas mixture of hydrogen and helium at solar-like abundance ratios ($\ce{H2}/\ce{He} = 0.85/0.15 \approx 5.7$; cf. $\approx 6.1$ in \citet{Asplund_2021_ChemicalMakeupSun}).
The effects of the background gas molecules on the transmission spectrum were included through their contributions to Rayleigh scattering and collision-induced absorption (CIA). For the latter, HITRAN data \citep{Karman_2019_UpdateHITRANCollisioninduced} were used, including CIA from \ce{H2}-\ce{H2} \citep[combined from][]{Abel_2011_CollisionInducedAbsorptionH2Pairs, Fletcher_2018_HydrogenDimersGiantplanet} and \ce{H2}-\ce{He} \citep{Abel_2012_InfraredAbsorptionCollisional} collisions.
The abundances of all molecular and atomic trace species are parametrized as logarithmic volume mixing ratios (VMRs), $X_\mathrm{x} = n_\mathrm{x}/n_\mathrm{total}$, and assumed to be vertically constant throughout the atmosphere (isochemical), a common simplification in atmospheric retrievals \citep[e.g.]{Madhusudhan_2019_ExoplanetaryAtmospheresKey, Lueber_2024_InformationContentJWST}.
Their contribution to the transmission spectrum was modeled using opacities from the ExoMol database \citep{Tennyson_2012_ExoMolMolecularLine} provided as correlated $k$-tables \citep{Chubb_2021_ExoMolOPDatabaseCross}. The used set of molecules included those previously detected for this planet, i.e. \ce{H2O} \citep[linelist:][]{Polyansky_2018_ExoMolMolecularLine}, \ce{CO2} \citep{Yurchenko_2020_ExoMolLineLists}, \ce{CO} \citep{Li_2015_RovibrationalLineLists}, and \ce{H2S} \citep{Azzam_2016_ExoMolMolecularLine}. Additionally, species that have been identified in JWST transmission spectra of other exoplanets (see Sect.~\ref{sec:introduction}), namely \ce{CH4} \citep{Yurchenko_2024_ExoMolLineLists}, \ce{SO2} \citep{Underwood_2016_ExoMolMolecularLine} and the potential nitrogen reservoirs \ce{NH3} \citep{Coles_2019_ExoMolMolecularLine}, and \ce{HCN} \citep{Barber_2014_ExoMolLineLists}, were included as well.
Finally, the potential impact of clouds was modeled using a non-gray cloud parametrization. It assumes a simple, homogeneous cloud structure, which effectively fills the atmosphere below a certain upper pressure limit $P_\mathrm{cloud}$ with cloud particles of constant mixing ratio $X_\mathrm{cloud}$, combined with a parametrized opacity $\kappa_\mathrm{cloud}$ for the cloud component,

\begin{equation}\label{eq:cloud_param}
     \kappa_\mathrm{cloud} = \frac{\kappa_0}{1 + \left( \frac{\lambda}{\lambda_0} \right)^{p} } \, ,
\end{equation}

which mimics the expected behavior of physical cloud particles by reaching a constant reference opacity $\kappa_0$ at short wavelengths $\lambda < \lambda_0$ and following a wavelength-dependent power-law of steepness $p$ at longer wavelengths, implying a cloud opacity in the optical that can be at most as steep as in the infrared and may flatten toward shorter wavelengths.
%This behavior resembles that of the parametrization from \citet{Lee_2013_AtmosphericRetrievalAnalysis}, while allowing for a slope deviating from the idealized case of Rayleigh scattering ($p=4$), similar to the extended parametrization from \citet{Kitzmann_2018_OpticalPropertiesPotential}.
% used in the WASP-39b information content study by \citep{Lueber_2024_InformationContentJWST}.

The retrievals were carried out using the ARCiS interface to the MultiNest code \citep{Feroz_2008_MultimodalNestedSampling, Feroz_2009_MULTINESTEfficientRobust, Feroz_2019_ImportanceNestedSampling}. Nested sampling was performed with 2500 live points and an efficiency rate of 0.3.
The planetary radius, the five temperature points, the chemical abundances of the included species, and the parameters of the cloud model were treated as free parameters. Since the cloud particle mixing ratio $X_\mathrm{cloud}$ and reference opacity $\kappa_{0}$ are exactly degenerate, only the former was varied, while the latter was arbitrarily fixed to $\SI{1000}{\centi\meter\squared\per\gram}$. The only quantity that can be constrained is the product of these two parameters. %$\kappa_{0}'$. % TODO? Think about turning this into an actual equation for improved readability?
The retrievals also allowed for a variable offset $\Delta\delta_\mathrm{0,Z}$ between observations from different instruments. The offsets were included as additional free parameters ($n-1$ parameters for $n$ instruments) sampled within the nested sampling framework.
In general, wide and uninformative priors were chosen (see Table~\ref{tab:retrieval_priors}). The prior for the spectral slope was specified to permit shapes ranging from flat to super-Rayleigh.

\begin{table}
\caption{Priors used in the retrievals conducted in this study.}\label{tab:retrieval_priors}
\input{tables/retrieval_priors} % TODO: Switch the R to a T when it comes o the temperature
\tablefoot{$\mathcal{U}$ = Uniform, $\mathcal{L}$ = Log-Uniform.}
\end{table}

Using the general setup described above, a number of different retrievals were performed. First, and forming the core of this study, seven retrievals were conducted using all possible combinations of the JWST instruments NIRISS, NIRSpec, and MIRI.
To specifically assess the benefits of the NIRISS SOSS mode, four retrievals were performed in which it was replaced by the overlapping but narrower and lower signal-to-noise (S/N) HST WFC3 mode.
Finally, to test the impact of the JWST data, especially NIRISS SOSS, on the retrieved slope in the optical, combined retrievals with different reductions of the STIS data were performed. For these retrievals, instrumental transit depth offsets were optimized separately using a least-squares minimization, following earlier work in this series \citep{Crouzet_2025_DetectionCO$_2$CO}, as including additional offset parameters directly in the nested sampling for the full STIS+JWST combinations would be computationally prohibitive. Tests for other multi-instrument combinations showed that this choice results only in minor differences in the resulting posterior (see Appendix~\ref{app_ssec:offset_treatment}).
Also, opacities of the atomic species sodium \citep[\ce{Na};][]{Allard_2019_NewStudyLine} and potassium \citep[\ce{K};][]{Allard_2016_KH2LineShapes} were additionally included.
To quantify detection significances of the included molecules, additional \enquote{leave-one-out} retrievals were conducted. Further, retrievals were performed using a gray instead of a non-gray cloud parametrization, and an isothermal in place of the standard $N$-point  $T$--$P$ profile, to assess the impact of these modeling assumptions. The influence of cloud opacity was also tested by fixing the spectral slope parameter to $p = 0$.

\section{Results}\label{sec:results}
\subsection{NIRISS data reduction}

\begin{figure*}
    \centering
    \includegraphics[width=\textwidth]{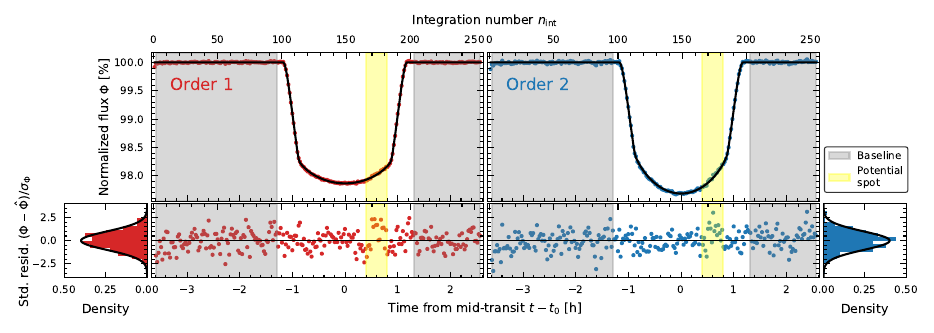}
    \caption{White lightcurves created from the first two spectral orders of the NIRISS SOSS data and combined best-fit model. The top panels show the observed data and fitted model lightcurves, while the bottom panels show the standardized residuals. The standardized residuals are the residuals divided by the uncertainty $\sigma_\Phi$ that combines the uncertainties in flux reported by the pipeline $\sigma_t$ with the additional jitter term $\sigma_\omega$ fitted by the \texttt{juliet} framework ($\sigma_\Phi = \sqrt{\sigma_t^2 + \sigma_\omega^2}$). The outer bottom panels compare the distribution of the standardized residuals to the shape expected for Gaussian noise. Shaded regions indicate the time intervals used for baseline normalization (gray) and a suspected spot crossing event (yellow).}\label{fig:white_lcurves}
\end{figure*}

The white lightcurve fits of the NIRISS SOSS data (see Fig.~\ref{fig:white_lcurves}) show generally good agreement between the data and the best-fit model ($\chi_{\nu}^{2} = 1.04$, $\mathrm{RMS} \, \text{(order 1/order 2)} = 114/189 \, \mathrm{ppm}$).
A localized deviation is observed in both lightcurves starting approximately 25 minutes after mid-transit and lasting for about 25 minutes ($n_\mathrm{int} \in [164, 179]$). This feature may be attributable to a spot crossing event. The impact of such events was explicitly tested using tweaked reductions (see Appendix~\ref{app_ssec:tweaks_description}). Differences in the resulting transmission spectrum were found to be minor (see Appendix~\ref{app_ssec:tweaks_impact}), and therefore the simpler spotless modeling was retained for the standard reduction.
While the one-dimensional distribution of standardized residuals closely follows the expected Gaussian shape ($p \gg 0.05$ for skewness and kurtosis tests), the time-resolved residuals still exhibit structured deviations.
The spectral lightcurves also generally show good agreement when accounting for the reduced S/N when compared to the white lightcurves ($\chi_{\nu}^{2} \sim 1.2 \text{--} 2.5$, $\mathrm{RMS} \sim 400 \text{--} 2000 \, \mathrm{ppm}$; see Fig~\ref{fig:spectral_lcurves} for representative example lightcurves).
While a direct comparison between the LDCs freely fitted for in the case of the white lightcurves and the fixed model values adopted for the spectral lightcurves is limited by their differing wavelength ranges, systematic differences are apparent. Specifically, the fitted LDCs tend to favor higher values of $u_1$ and lower values of $u_2$ than the corresponding model values. These differences were further investigated using additional tweaked reductions (see Appendix~\ref{app_ssec:tweaks_description} and Fig.~\ref{fig:limb_darkening}).
Accounting for an offset of $250$ ppm in the transit depth (well within the adopted prior range of $1000$ ppm), which was treated as a free parameter in the retrievals, brings the NIRISS transmission spectrum into good agreement with the HST WFC3 spectrum at overlapping wavelengths (see Fig.~\ref{fig:transspec_comparison}).

\subsection{Retrievals}
\begin{figure*}[h]
    \centering
    \includegraphics[width=\textwidth]{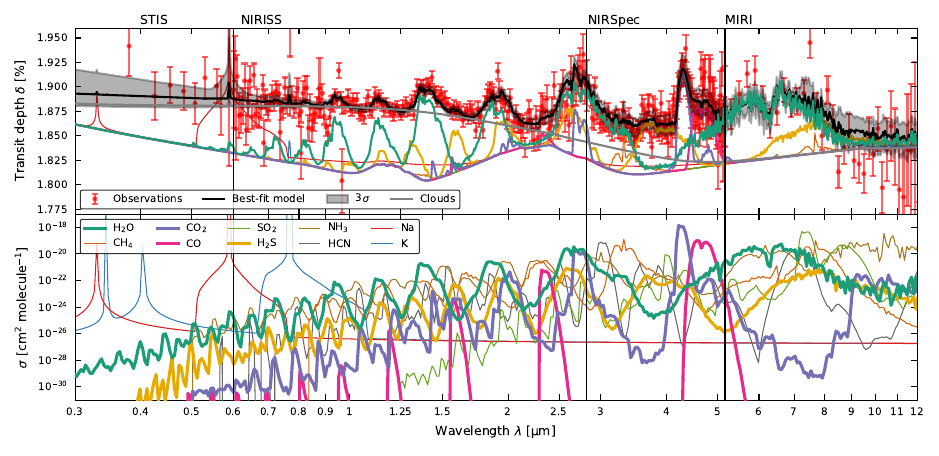}
    \caption{Visualization of the HAT-P-12b transmission spectrum and the underlying opacity contributions.
    \textit{Top:} Combined spectra from the three JWST instruments and HST STIS \citep{Alexoudi_2020_RoleImpactParameter} with $1\sigma$ uncertainties, %(red markers),
    together with the best-fit model %(black line)
    with its $3\sigma$ credible interval.%(gray shading).
    Model spectra including only the opacity of individual contributors (chemical species or non-gray cloud parametrization) are also shown. 
    \textit{Bottom:} Absorption cross sections $\sigma$ of all tested chemical species; significantly detected species are emphasized with thicker lines.}\label{fig:model+opacities}
\end{figure*}

To interpret the retrieval results, individual molecular bands were identified using the database by \citet{Crovisier_2002_JaquesCrovisierMolecular}. The corresponding opacity contributions of the different species are shown in Figure~\ref{fig:model+opacities}.
Frequentist $n_\sigma$ detection significances were derived from differences in Bayesian log-evidence $\Delta \ln Z$ using the method from  \citet{Sellke_2001_CalibrationValuesTesting, Trotta_2008_BayesSkyBayesian}. While this approach is commonly used in atmospheric retrieval studies \citep[e.g.][]{Benneke_2013_HowDistinguishCloudy, Kreidberg_2015_DetectionWaterTransmission, Edwards_2023_ExploringAbilityHubble}%Ahrer_2022_EarlyReleaseScience
, the obtained detection significances should be treated with caution \citep{Kipping_2025_ExoplaneteersKeepOverestimating, Welbanks_2025_ChallengesDetectingGases}.
The results from the three sets of retrievals are discussed separately in the following sections.

\subsubsection{JWST instrument combinations}

\begin{figure}
    \centering
    \includegraphics[width=\columnwidth]{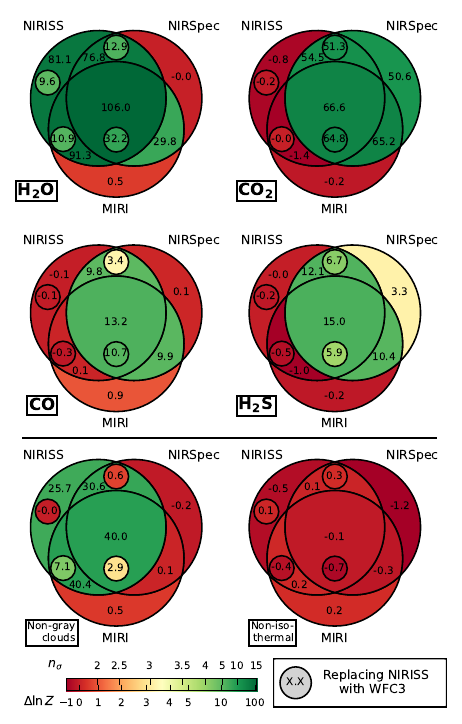}
    \caption{Venn diagrams of JWST instrument combinations showing detection significances of the four significantly detected molecules (top) and the preference for two modeling assumptions (bottom). Additional circles were used to show the significances obtained when replacing the JWST NIRISS SOSS with the HST WFC3 instrument data. The significances are reported as log-evidence differences $\Delta \ln{Z}$ and visualized using a custom non-linear color scale. The colorbar also provides a conversion to $n_\sigma$ detection significances following the calibration scale from \citet{Benneke_2013_HowDistinguishCloudy}.}\label{fig:signif_jwst_wfc3}
\end{figure}

Significant detections ($> 3\sigma$) are achieved for four molecules: \ce{H2O}, \ce{CO2}, \ce{CO}, and \ce{H2S} (see Fig.~\ref{fig:signif_jwst_wfc3}). These are the same species found in \citet{Crouzet_2025_DetectionCO$_2$CO}.
Any combination of instruments involving NIRISS lead to a highly significant detection of \ce{H2O}. Even for NIRISS alone the significance is above $12 \sigma$. The reason for this is the coverage of numerous dominant and well isolated absorption bands.
A detection without NIRISS requires combining NIRSpec and MIRI. NIRSpec captures only the long-wavelength tail of the fundamental stretching bands ($\nu_1$ and $\nu_3$), with the rest of its range dominated by other molecular features. In contrast, MIRI is limited by lower S/N, a consequence of the declining stellar flux in the mid-infrared.
This combination yields a detection significance of $\sim 8 \sigma$, which drastically increases to $\sim 15 \sigma$ when NIRISS is included as well. 
A significant detection of \ce{CO2} is only possible when NIRSpec data is included, yielding very high significances of $>10 \sigma$. This is due to the strong and well-isolated asymmetric stretch band $\nu_3$ at \SI{\sim 4.3}{\micro\meter} falling within NIRSpec's range.
For \ce{CO}, instrument combinations that include NIRSpec reach $\sim 5 \sigma$, probably due to the fundamental band at \SI{\sim 4.7}{\micro\meter}. Notably, NIRSpec alone does not enable a significant \ce{CO} detection, as the retrieval can match the observed opacity using only \ce{H2O} and \ce{CO2}. 
However, as shown by \citet{Crouzet_2025_DetectionCO$_2$CO}, the alternative TEATRO reduction can produce substantially higher detection significances, indicating that the result remains sensitive to the data reduction approach.
A significant detection of \ce{H2S} is only possible when NIRSpec data are included, due to its symmetric stretch band $\nu_1$ near \SI{3.8}{\micro\meter}, which lies in a spectral region where \ce{H2O} and \ce{CO2} opacity reaches a local minimum. Using NIRSpec alone yields a lower significance of $\sim 3.1 \sigma$, while combinations with other instruments generally achieve $>4.9 \sigma$. The \ce{H2S} opacity in this region could also be mimicked by gray cloud opacity, but added constraints from NIRISS and MIRI help break this degeneracy, enabling a more robust detection. This effect was also observed in \citet{Crouzet_2025_DetectionCO$_2$CO}, where \ce{H2S} was only detected for the CASCADe reduction. To test whether the TEATRO reduction also yields a detection of \ce{H2S} when combined with the other JWST data, an additional retrieval was performed. The resulting detection significance is lower at $\sim 3.7 \sigma$, but remains significant.

\begin{figure}
    \centering
    \includegraphics[width=\columnwidth]{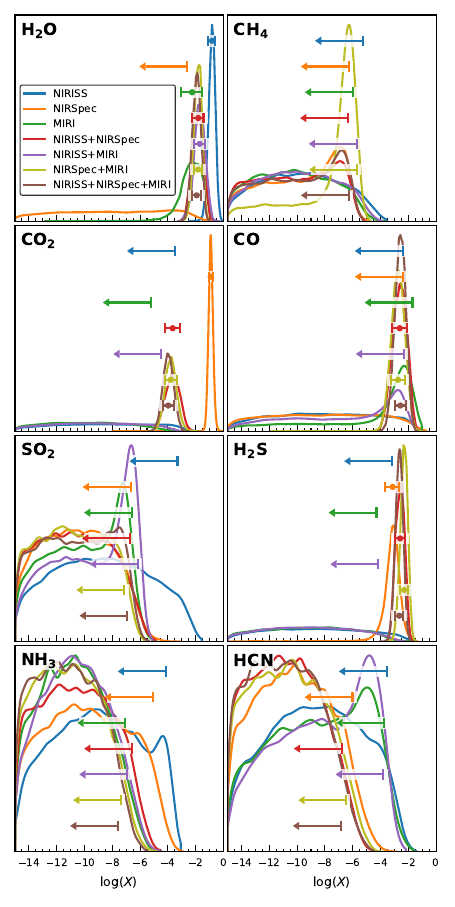}
    \caption{Marginal posterior distributions of the chemical abundances for all included molecules, retrieved using different JWST instrument combinations. The shown distributions were derived from the samples via kernel density estimation (KDE; Gaussian kernel with bandwidth $h=0.1
    $). For each distribution, markers indicate either the $1\sigma$ credible interval, in cases of well-constrained posteriors, or the 95th percentile upper limit (arrows) when only an upper bound can be established.}\label{fig:marg_abunds_jwst}
\end{figure}

Significant detections correspond to narrow, well-constrained posteriors, whereas non-detections yield broad distributions with only upper limits and extended tails toward lower abundances (except for \ce{H2O} with MIRI; see Fig.~\ref{fig:marg_abunds_jwst} and Table~\ref{tab:marg_abunds_jwst}).
For molecules with significant detections, the posteriors are mostly consistent within their $1\sigma$ intervals. Combining several instruments often lead to consistent but more tightly constrained abundances compared to single instruments. Notable outliers are the abundance constraints for \ce{H2O} with NIRISS and \ce{CO2} with NIRSpec, both yielding significantly higher values. This underscores the need for broad wavelength coverage to constrain the atmospheric structure, since retrievals based on limited data can yield precise but biased posteriors.
For molecules without significant detections, combining instruments still proves beneficial, typically leading to tighter 95th percentile upper limits on their abundances.
For \ce{CH4}, most upper limits cluster around $-6$ dex, with slightly higher values for the NIRISS and NIRSpec+MIRI retrievals, the latter even showing a pronounced peak at the upper bound.
For \ce{SO2}, the upper limits fall into distinct tiers. The NIRISS-only retrieval yields the weakest constraint at approximately $-3.3$ dex, approximately $3$ dex higher than all others. This is not surprising since both NIRSpec and MIRI have regions with dominant \ce{SO2} absorption bands while NIRISS does not. Retrievals including only MIRI or NIRISS+MIRI exhibit peaks at the upper limit, but the peak disappears when more instruments are included. The most stringent limits are obtained when both NIRSpec and MIRI are included (either with or without NIRISS), lowering the upper limit by $\sim 0.2$ dex compared to other combinations.
For \ce{NH3}, upper limits again generally tighten as more instruments are added. MIRI has the largest impact, probably due to strong, broad opacity beyond \SI{\sim 9}{\micro\meter} from the molecule's symmetric bend \enquote{umbrella} mode $\nu_2$, followed by NIRSpec, while NIRISS contributes the least.
For \ce{HCN}, two groups of constraints can be distinguished. Retrievals that exclude NIRSpec yield significantly weaker limits. Including NIRSpec lowers the upper limit by more than $2.2$ dex, most likely because its wavelength range covers both the fundamental \ce{C}--\ce{H} stretch $\nu_3$ at \SI{\sim 3.0}{\micro\meter} and the combination band of \ce{C}--\ce{N} stretch and the bend mode $\nu_1 + \nu_2$ at \SI{\sim 3.6}{\micro\meter}.

\begin{figure}
    \centering
    \includegraphics[width=\columnwidth]{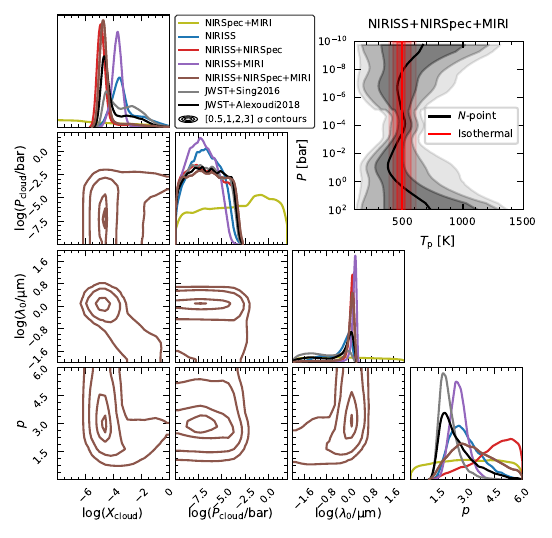}
    \caption{\textit{Left:} Excerpt from corner plot focusing on parameters of the cloud parametrization.
    %The off-diagonal subplots display the two-dimensional joint posterior distributions, highlighting potential parameter correlations. 
    The joint posterior distributions (off-diagonal panels) are only shown for the retrieval using the data from all JWST instruments and are visualized using contour lines smoothed with a Gaussian kernel ($\sigma=1$).
    %The diagonal panels show the one-dimensional marginal distributions of the different parameters. 
    The marginal posterior distributions (diagonal panels) are additionally shown for several exemplary JWST instrument combinations, as well as the results from the retrievals under the inclusion of the two different HST STIS reductions. The same KDE as in Figure~\ref{fig:marg_abunds_jwst} was applied. 
    \textit{Upper Right:} Comparison of $T$--$P$ profiles derived for more complex $N$-point and simpler isothermal parametrization for the case of the three JWST instrument combination. Lines indicate the median value, while shaded regions are used to highlight the [0.5, 1,2,3] $\sigma$ credible intervals.}\label{fig:corner_clouds+tp_limits}
\end{figure}

In addition to the tests on molecular detection significance, significance tests of key atmospheric assumptions, such as the choice between sloped non-gray and flat gray cloud opacities, were also performed (see Fig.~\ref{fig:signif_jwst_wfc3}).
A preference for the sloped model is only significant, at more than $\sim 7.5 \sigma$, when the NIRISS data is included. This is expected, since the effect of the spectral slope becomes more pronounced at short wavelengths. Additionally, there is a dip in transit spectrum around \SI{2.3}{\micro\meter} between two \ce{H2O} bands which cannot be adequately reproduced by a gray cloud model (see Fig.~\ref{fig:model+opacities}). Such a feature was also observed in \citet{Feinstein_2023_EarlyReleaseScience} where it was also interpreted as an indication of a non-gray cloud opacity and in their case even inhomogeneous cloud coverage.
Figure~\ref{fig:corner_clouds+tp_limits} shows marginal and joint posteriors of the cloud parameters for select instrument combinations. The joint distributions reveal notable correlations between several parameters. Retrievals using only NIRSpec and MIRI, whether individually or combined, yield flat posteriors matching the priors. In contrast, combinations including NIRISS provide clear constraints. The cloud particle mixing ratio $X_\mathrm{cloud}$ is constrained to be above $-4.4$ dex. The cloud-top pressure $P_\mathrm{cloud}$ is generally constrained to be lower than \SI[parse-numbers = false]{10^{-3.6}}{\bar}, requiring the cloud to extend into the probed layers. The reference wavelength $\lambda_0$ is limited to be below \SI{2}{\micro\meter} for NIRISS+MIRI and below \SI{1.5}{\micro\meter} for the remaining combinations, matching the shape of the NIRISS spectrum at short wavelengths. Finally, all NIRISS retrievals also favor sub-Rayleigh slopes with $p < 4$, except for NIRISS+NIRSpec, where a super-Rayleigh slope with $p > 4$ is preferred above the 32nd percentile.
No significant preference for the more complex $N$-point profile over the isothermal temperature profile was found.
The information content of the data therefore does not justify the added complexity introduced by a non-isothermal profile.
For the combination of all three instruments, the use of a complex profile has negligible impact in the pressure range between \SIrange{1e-5}{1e-3}{\bar} (see Fig.~\ref{fig:corner_clouds+tp_limits}), which is the region typically probed by transmission spectroscopy. Outside this range, the median $N$-point profile deviates noticeably from the isothermal profile, but remains poorly constrained.

Overall, these results might give the impression that MIRI generally provides little additional benefit compared to NIRISS and NIRSpec. However, this largely reflects the characteristics of this particular planet, as it does not exhibit any strong MIRI-specific features \citep[e.g. \ce{SO2} and silicate clouds;][]{Dyrek_2024_SO2SilicateClouds}.

\subsubsection{NIRISS vs. WFC3} % TODO? Perhaps use a less informal title

The retrievals using HST WFC3 instead of NIRISS SOSS show that most molecules significantly detected with NIRISS are also detected with WFC3 (see Fig.~\ref{fig:signif_jwst_wfc3}). A notable exception is \ce{CO}, for which a significant detection only occurs when WFC3 is combined with NIRSpec and MIRI. This is probably due to the limited wavelength coverage of WFC3, which excludes the strongest \ce{CO} absorption band in the NIRISS range, the first overtone ($\Delta\nu = 2$) near \SI{2.3}{\micro\meter}.
To quantify changes in the marginal posterior distributions, we calculate the difference between the median values for bounded constraints or the upper 95th percentile for upper limits. These are reported both in dex and in units of $\sigma$, where $\sigma$ is the combined $1\sigma$ uncertainty of the two distributions, obtained by adding their individual $1\sigma$ uncertainties in quadrature (dex|$\sigma$ shorthand is used when values in dex and $\sigma$ are identical).
Differences in the retrieved chemical abundances are generally small for multi-instrument combinations ($\lesssim 0.7$~dex|$\sigma$; see Fig.~\ref{fig:marg_abunds_wfc3}), a notable outlier being the \ce{CO2} abundance for WFC3+NIRSpec ($1.4$~dex, $1.7 \sigma$).
%The high \ce{CO2} abundance was not found in the corresponding retrieval by \citet{Crouzet_2025_DetectionCO$_2$CO}, most likely due to the use of a gray rather than a non-gray cloud parametrization.
The difference becomes negligible ($0.03$~dex, $0.05\sigma$) once MIRI is included as well.
For single-instrument retrievals, however, the results can be substantially more discrepant. \ce{H2O} is the only molecule significantly detected with WFC3 alone; instead of a narrow, well-constrained posterior, a distinctly different ($2.5$~dex, $1.9 \sigma$) bimodal distribution is retrieved. For most other molecules, except for \ce{CO2} and \ce{CO} ($\lesssim 0.28$~dex, $\lesssim 0.05 \sigma$), the upper limits differ considerably between the single-instrument retrievals ($\lesssim 2.8$~dex, $\lesssim 0.6 \sigma$). For \ce{CH4}, \ce{H2S}, \ce{NH3}, and \ce{HCN}, WFC3 yields lower upper abundance limits than NIRISS, highlighting the risk of overconfident interpretations from WFC3 data alone. 

In the case of HAT-P-12b, both WFC3- and NIRISS-only retrievals attribute their common spectral features to water. However, the WFC3 G141 grism covers only a single full \ce{H2O} band, making it susceptible to confusion with other molecules, particularly with \ce{CH4}, which has a similar although narrower and slightly shifted feature. An example for this confusion is K2-18b (\ce{H2O}: \citealt{Benneke_2019_WaterVaporClouds, Tsiaras_2019_WaterVapourAtmosphere}; \ce{CH4}: \citealt{Bezard_2022_MethaneDominantAbsorber, Madhusudhan_2023_CarbonbearingMoleculesPossible}).
%, where the original WFC3 spectrum was interpreted as showing a \ce{H2O} feature \citep{Benneke_2019_WaterVaporClouds}, but later JWST observations (NIRISS+NIRSpec) identified the feature as \ce{CH4} instead \citep{Madhusudhan_2023_CarbonbearingMoleculesPossible}, enabled by the broader spectral coverage across multiple bands.

While differences in molecular detections and abundance constraints are mostly negligible when several instruments are combined, NIRISS and WFC3 markedly differ in their ability to constrain the cloud opacity. With NIRISS, a strong preference for non-gray clouds is found both in the single instrument retrieval and in any combination with other instruments ($\gtrsim 7.5\sigma$). In contrast, WFC3 alone does not yield significant evidence. Only the WFC3+MIRI combination, a combination with a significant \SI{3}{\micro\meter} wavelength gap, shows a clear preference ($4.2\sigma$), which weakens to tentative levels ($2.9 \sigma$) once NIRSpec is also included.
%A significant preference for a non-gray cloud opacity is only found when MIRI is included, illustrating the limited diagnostic power of the WFC3 wavelength range. Even when there is a significant preference, the significance is lower with WFC3 ($\lesssim 4.6\sigma$) than with NIRISS ($\gtrsim 7.5\sigma$).

\subsubsection{Inclusion of STIS data}\label{ssec:inclusion_stis}

By combining the disparate STIS spectra from \citet{Sing_2016_ContinuumClearCloudy} (hereafter \enquote{Sing2016}) and \citet{Alexoudi_2018_DecipheringAtmosphereHATP12b} (hereafter \enquote{Alexoudi2018}) in separate, dedicated retrievals with the JWST data, especially NIRISS, it can be tested whether the retrieved steepness of the slope is still affected by differences between the STIS datasets.
Given the dominance of the cloud opacity at optical wavelengths, it is unsurprising that the retrieved chemical abundances are only marginally affected ($\lesssim 0.27$~dex|$\sigma$) by the inclusion of STIS data.
Neither with nor without STIS the additionally included atomic species of sodium or potassium get detected. For potassium, the retrievals produce very similar posteriors, with 95th percentile upper limits around $-6.6 \pm 0.1$ dex.
Sodium shows slightly higher limits without STIS, peaking near $-2.7$ dex, and lower limits with STIS around $-3.75 \pm 0.25$ dex.

The cloud parametrization is more strongly impacted (see Fig.~\ref{fig:corner_clouds+tp_limits}). All parameters, except for the cloud-top pressure $P_\mathrm{cloud}$, are significantly affected. 
In particular, the retrieved spectral slope $p$ shifts to lower values when STIS data are included alongside the JWST observations (JWST+Sing2016: $1.9_{-0.4}^{+0.6}$; JWST+Alexoudi2018: $2.3_{-0.6}^{+1.2}$) compared to the retrieval using JWST data alone ($3.3_{-1.1}^{+1.4}$). A preference for sub-Rayleigh slopes ($p<4$), i.e. up to the upper uncertainty bounds, is therefore only obtained when STIS data are included, while differences between the two STIS datasets themselves have little impact.
It should be noted, however, that the adopted cloud parametrization does not allow for an optical slope steeper than at longer infrared wavelengths. This coupling may limit the ability of the retrievals to reproduce Rayleigh-like scattering behavior in the optical.
Nevertheless, both STIS-inclusive cases show a stronger preference for sub-Rayleigh slopes than the JWST-only retrieval.
 
\section{Discussion}\label{sec:discussion}
\subsection{Elemental composition}
Molecular abundances inferred from transmission spectra can be used to estimate elemental abundance ratios, such as the carbon-to-oxygen (C/O) ratio and metallicity.
%The interest in characterizing these ratios arises from the comparisons to the solar system mass-metallicity trend \citep{Thorngren_2016_MassMetallicityRelationGiant, Welbanks_2019_MassMetallicityTrendsTransiting} and suggested links between the atmospheric and interior composition \citep{Thorngren_2019_ConnectingGiantPlanet}. Extending this chain of inference, it has been suggested that these values could trace planetary formation \citep{Molliere_2022_InterpretingAtmosphericComposition, Khorshid_2024_ConstrainingFormationWASP39b} by constraining the formation location in the disk \citep{Oberg_2011_EffectsSnowlinesPlanetary}, the planet's migration history \citep{Mordasini_2016_ImprintExoplanetFormation}, and the formation mechanism itself \citep{Madhusudhan_2017_AtmosphericSignaturesGiant}.
These quantities are of particular interest because they can be compared to solar system mass–metallicity trends \citep{Thorngren_2016_MassMetallicityRelationGiant, Welbanks_2019_MassMetallicityTrendsTransiting} and may provide insights into planetary formation and migration processes \citep{Oberg_2011_EffectsSnowlinesPlanetary, Mordasini_2016_ImprintExoplanetFormation, Thorngren_2019_ConnectingGiantPlanet, Molliere_2022_InterpretingAtmosphericComposition, Khorshid_2024_ConstrainingFormationWASP39b}.

\begin{figure}
    \centering
    \includegraphics[width=0.5\textwidth]{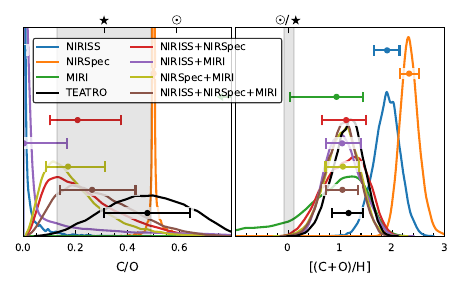}
    \caption{Marginal posterior distributions of the C/O ratio and metallicity ([(C+O)/H]]) derived from the abundances of the included molecular species for all JWST instrument combinations.
    An additional retrieval of all JWST instruments using the NIRSpec TEATRO instead of the standard CASCADe reduction is also shown (labeled TEATRO).
    The interval markers follow the convention introduced in Figure~\ref{fig:marg_abunds_jwst}. Solar \citep[$\odot$;][]{Asplund_2021_ChemicalMakeupSun} and stellar \citep[$\star$;][]{Biazzo_2022_GAPSProgrammeTNG} values are indicated on the top axis.
    Gray shading marks the uncertainty in the stellar values.}\label{fig:bulk_comp_jwst}
\end{figure}

The weighted retrieval samples were used to derive C/O ratios and metallicities for the various JWST instrument combinations. All molecules, detected or not, were included in the calculation, such that undetected molecules still contributed to the overall uncertainty. For the metallicity $Z$, the ratio of carbon and oxygen to hydrogen was used as a proxy, since the other two atomic species, sulfur and nitrogen, are not as reliably constrained.
All retrievals generally favor super-solar metallicities (see Fig.~\ref{fig:bulk_comp_jwst}). Single instrument retrievals, however, prove less reliable. For NIRISS and NIRSpec, the aforementioned overestimation of \ce{H2O} or \ce{CO2} results in metallicities around $\sim 100 \times$ solar, well above the $\sim 10\times$ solar inferred from most combination retrievals.

Given the strong influence of the NIRSpec data on the inferred abundances of carbon-bearing species, an additional retrieval was performed using the full JWST dataset, replacing the CASCADe NIRSpec reduction with the TEATRO reduction \citep{Crouzet_2025_DetectionCO$_2$CO}.
The impact on metallicity is minimal, the C/O ratio, however, is more strongly impacted (values given below).
This can be attributed to shifts in the molecular abundances. While these all stay within their $1\sigma$ uncertainty intervals, the ratio of $\ce{H2O}/\ce{CO}$ drops by more than half, while $\ce{CO}/\ce{CO2}$ nearly doubles. The retrievals based on CASCADe yield C/O ratios broadly consistent with the stellar value, with a preference for sub-stellar values. In contrast, the TEATRO-based retrieval favors a wider range from stellar to solar values, a difference already noted in \citet{Crouzet_2025_DetectionCO$_2$CO} although using the WFC3+NIRSpec combination instead of NIRISS+NIRSpec+MIRI.

The values inferred from the retrievals including all JWST instruments are $\mathrm{C}/\mathrm{O} = 0.26_{-0.12}^{+0.17}$ and $Z = 11_{-6}^{+11}$ for CASCADe, and $\mathrm{C}/\mathrm{O} = 0.48_{-0.17}^{+0.16}$ and $Z = 15_{-8}^{+13}$ for TEATRO. 
These results are broadly consistent with previous results from \citet{Crouzet_2025_DetectionCO$_2$CO}, despite differences in the retrieval setups.
Pre-JWST results based on HST, Spitzer, and ground-based photometry and made under the assumption of equilibrium chemistry typically found near-solar C/O ratios, but are still compatible with sub-solar values \citep{Wong_2020_OpticalNearinfraredTransmission, Yan_2020_LBTTransmissionSpectroscopy, Panek_2023_ReanalysisEquilibriumChemistry}. Metallicities varied widely, from notably lower values \citep[$<10\times$ solar;][]{Yan_2020_LBTTransmissionSpectroscopy, Panek_2023_ReanalysisEquilibriumChemistry} to much higher estimates \citep[$>100\times$ solar;][]{Wong_2020_OpticalNearinfraredTransmission, Jiang_2021_EvidenceStellarContamination}. These comparisons underscore the importance of JWST’s broad spectral coverage and precision for obtaining robust, though potentially still reduction- and model-dependent, estimates of elemental abundances.

\subsection{Behavior in the optical}\label{subsec:behavior_optical}

As discussed in detail in the introduction, there have been discrepant results regarding the shape of the transmission spectrum of HAT-P-12b at optical wavelengths.
Combining either of the disparate STIS data reductions, with the JWST data from all three instruments, result in a clear preference of a low steepness slope, which is very different from the results obtained when analyzing the STIS data on its own \citep{Alexoudi_2018_DecipheringAtmosphereHATP12b}.
As discussed in Sections \ref{sec:retrievals} and \ref{ssec:inclusion_stis}, the adopted cloud parametrization intrinsically limits the optical cloud opacity to be at most as steep as at longer infrared wavelengths. While this physically motivated assumption provides a compact description across the full wavelength range, alternative cloud opacity treatments will be explored for the JWST data in future work (Bouwman et al., in prep.).
The preference for a shallow optical slope may be further reinforced by the greater weight of the JWST data in the retrievals, as their lower uncertainties and larger number of datapoints compared to the STIS data lead the sampling algorithm to favor them.
Given the importance of the NIRISS data when it comes to constraining the shape of the spectral slope, as illustrated by its necessity to obtain a significant preference for the non-gray cloud parametrization (see Fig.~\ref{fig:signif_jwst_wfc3}), ensuring its reliability, particularly at its short wavelength end, remains essential.
Contributing to this as a potential source of bias is stellar contamination from spots and faculae \citep{Rackham_2018_TransitLightSource}. While the impact of a potential spot crossing event was tested (see Appendix \ref{app_ssec:tweaks_impact}), no attempt was made to characterize unocculted stellar heterogeneities. 
While HAT-P-12 has been characterized as a low activity host star based on a low estimated rotation rate \citep{Hartman_2009_HATP12bLowDensitySubSaturn, Mancini_2018_GAPSProgrammeHARPSN} and low variability in photometric measurements \citep{Mancini_2018_GAPSProgrammeHARPSN, Wong_2020_OpticalNearinfraredTransmission, Yan_2020_LBTTransmissionSpectroscopy}
, differences observed in ground-based optical photometry of HAT-P-12b at two separate visits have been interpreted as evidence of unocculted stellar heterogeneities \citep{Jiang_2021_EvidenceStellarContamination}. Their impact is not quantified during the lightcurve fitting, but is instead typically assessed as part of the atmospheric retrieval \citep{Pinhas_2018_RetrievalPlanetaryStellar}. Although such an analysis is beyond the scope of this study, an analysis similar to the one presented in \citet{Fournier-Tondreau_2024_NearInfraredTransmissionSpectroscopy} could be conducted in the future.

\subsection{Comparison to information content of WASP-39b}

Understanding how results from a single target, such as HAT-P-12b, generalize to other exoplanets is key to characterizing the broader information content of JWST transmission spectra. 
As mentioned in the introduction, one of the first exoplanets to be characterized in detail with most of the applicable near-infrared JWST modes, including NIRISS SOSS \citep{Feinstein_2023_EarlyReleaseScience} and NIRSpec G395H \citep{Alderson_2023_EarlyReleaseScience},
was WASP-39b.
Compared to HAT-P-12b, it is hotter ($T_\mathrm{eq}$: $+\SI{211 \pm 11}{\kelvin}$), larger ($R_\mathrm{p}$: $\SI{139 \pm 6} {\percent}$), and less dense ($\rho_\mathrm{p}$: $\SI{69 \pm 11}{\percent}$) \citep{Mancini_2018_GAPSProgrammeHARPSN}. Information content studies conducted on WASP-39b considered either only single-instrument retrievals \citep{Lueber_2024_InformationContentJWST, Schleich_2024_KnobsDialsRetrieving}, or combined retrievals with HST WFC3 \citep{Constantinou_2023_EarlyInsightsAtmospheric} or HST STIS \citep{Fisher_2024_JWSTNIRISSHST}. None of these studies included the MIRI LRS data obtained as part of another program \citep{Powell_2024_SulfurDioxideMidinfrared}.

Significant molecular detections for single-instrument retrievals of matching modes are generally consistent between the two planets, one notable difference being a detection of potassium with NIRISS SOSS in the case of WASP-39b \citep{Lueber_2024_InformationContentJWST, Fisher_2024_JWSTNIRISSHST}.
The detection of \ce{H2S} for WASP-39b was found to be sensitive to small differences between data reductions, but as for HAT-P-12b it was found that combined retrievals with data at other wavelengths can lead to a reduction-independent detection \citep{Constantinou_2023_EarlyInsightsAtmospheric}.
In contrast, retrievals using only NIRSpec G395H reveal at most weak and model-dependent evidence \citep{Lueber_2024_InformationContentJWST}.
This suggests more generally that the molecule's $\nu_1 + \nu_3$ combination band near $\SI{3.9}{\micro\meter}$ provides sensitivity to its presence, but that a robust detection requires complementary wavelength coverage.

When comparing retrieved abundances, differences between data reductions in single-instrument retrievals of WASP-39b led to variations of up to 1 dex \citep{Constantinou_2023_EarlyInsightsAtmospheric}.
No equivalent tests were conducted for HAT-P-12b \citep[but see][for NIRSpec]{Crouzet_2025_DetectionCO$_2$CO}, although single-instrument retrievals yielded abundance differences of $\sim 1$ dex or more for select molecules. For both planets, combining multiple instruments mitigated these discrepancies, highlighting the risk for overconfident results from single-instrument retrievals.
Consistent with this trend, abundance differences between retrievals of HAT-P-12b arising from different reductions of individual instruments, when included in combined retrievals with the other JWST instruments, lead to much smaller differences 
(NIRISS: $\lesssim 0.3$~dex, $\lesssim 0.11 \sigma$ except for \ce{H2S} $\lesssim 0.5 \sigma$; NIRSpec: $\lesssim 0.5$~dex, $\lesssim 0.6 \sigma$ except for \ce{H2S} $\lesssim 1.1 \sigma$).
The use of HST WFC3 instead of JWST NIRISS in single-instrument retrievals leads to less reliable constraints for both planets \citep{Fisher_2024_JWSTNIRISSHST}. 
While for WASP-39b, the inclusion of STIS data in NIRISS retrievals led to shifts of up to $\sim 1$ dex in the retrieved \ce{H2O} abundance \citep{Fisher_2024_JWSTNIRISSHST}, adding STIS to the combined JWST retrievals of HAT-P-12b produced no discernible change in the derived chemical abundances. This further highlights the improved consistency achieved when jointly fitting multiple JWST datasets.

Although the specific parametrizations differ, the same two modeling assumptions and scenarios were tested for the two planets. As for HAT-P-12b, NIRISS SOSS led to a significant preference for non-gray cloud behavior for WASP-39b. Consistent with this study, it was found that HST WFC3 is insufficient to obtain this preference \citep{Fisher_2024_JWSTNIRISSHST}. Contrary to HAT-P-12b such a preference could, however, also be obtained for single-instrument retrievals of NIRSpec G395H in the case of WASP-39b. The preference was attributed to the cloud opacity being able to compensate for molecular opacity from \ce{H2O} at the instrument's short wavelength end \citep{Lueber_2024_InformationContentJWST}. The other modeling assumption tested for both planets is the use of an isothermal or a more complex $T$--$P$ profile. For WASP-39b a preference for a non-isothermal profile was obtained for various single-instrument retrievals \citep{Lueber_2024_InformationContentJWST, Fisher_2024_JWSTNIRISSHST}, while for HAT-P-12b no such preference was found for any of the instrument combinations. This could be indicative of differences in the shape of the $T$--$P$ profile between the two planets in the pressure range probed by transmission spectroscopy. \citet{Schleich_2024_KnobsDialsRetrieving} used simulated but observation-based NIRSpec PRISM model spectra of WASP-39b to show that oversimplified $T$--$P$ profiles can bias retrieved abundances. While it is not possible to compare retrieved to actual abundances for real observations, differences are generally $<0.3$~dex|$\sigma$, with most retrievals yielding slightly higher abundances under the $N$-point profile (see Fig~\ref{fig:comp_abund_isothermal}). Exceptions occur mostly in single-instrument retrievals, 
especially NIRISS and to a lesser extent NIRSpec alone. The largest outlier is the NIRISS \ce{H2O} abundance, which is bimodal under the isothermal profile due to degeneracies between planetary radius and cloud parameters, and overconfidently constrained to the higher abundance mode under the $N$-point profile.

\section{Conclusions}\label{sec:conclusions}

This study presents a JWST NIRISS SOSS transmission spectrum of the warm sub-Saturn HAT-P-12b.
%The dataset was carefully analyzed and its robustness tested by applying a series of point-wise adjustments throughout the data reduction process.
Combined with additional JWST observations obtained with NIRSpec G395M \citep{Crouzet_2025_DetectionCO$_2$CO} and MIRI LRS (Bouwman et~al., in prep.), the full wavelength range accessible with JWST (excluding MIRI MRS) is covered.
This combined dataset, supplemented by archival HST data where relevant, was used for a detailed information content study via atmospheric retrievals.
Key findings from the data reduction and retrieval analysis are:

\begin{enumerate}
    \item \textbf{Molecular constraints} \\
    The retrievals resulted in significant detections of four molecules: \ce{H2O}, \ce{CO2}, \ce{CO}, and \ce{H2S}. All were previously reported in \citet{Crouzet_2025_DetectionCO$_2$CO}, although \ce{H2S} was detected only tentatively with one of the two NIRSpec reductions. In combined retrievals using all JWST modes, \ce{H2S} was significantly detected for both reductions. Instrument-dependent detection significances can be well explained by the presence or absence of distinct molecular features within the covered wavelength range. For all molecules, except \ce{H2O}, NIRSpec was required for detection, but not necessarily sufficient (\ce{CO}, \ce{H2S}). The abundances of significantly detected molecules can be highly discrepant in single instrument retrievals (\ce{H2O}: NIRISS, \ce{CO2}: NIRSpec), but are mostly consistent in multi-instrument retrievals. For non-detected molecules (\ce{CH4}, \ce{SO2}, \ce{NH3}, \ce{HCN}), combined retrievals can lead to improved upper limits.
    
    \item \textbf{NIRISS vs. WFC3} \\
    Replacing NIRISS with WFC3 reduced molecular detection significances, though they generally remained significant.
    Abundance estimates were mostly consistent, but showed discrepancies in single-instrument retrievals (\ce{H2O}) and occasionally in two-instrument combinations (\ce{CO2} with WFC3+NIRSpec).
    A key limitation of WFC3 is its weaker ability to constrain the cloud opacity. While NIRISS consistently indicated non-gray behavior, WFC3 yielded reliable evidence only in the WFC3+MIRI combination.
    
    \item \textbf{Elemental abundance ratios} \\
    Elemental abundance ratios were derived for each instrument combination. The sensitivity of three out of four detected molecules (\ce{CO2}, \ce{CO} \& \ce{H2S}) to the NIRSpec data strongly influences the results. Metallicities are discrepant in single-instrument retrievals, but converge to median values around 11–15$\times$ solar in other combinations. Constraining the C/O ratio requires the NIRSpec data. A general, though weak, preference for sub-stellar values disappears when replacing the NIRSpec CASCADe reduction with TEATRO, highlighting the need for caution when interpreting abundance ratios.

    \item \textbf{Behavior in the optical} \\
    Combining the JWST data with STIS observations yields a moderate scattering slope with a steepness below that expected for Rayleigh scattering regardless of the chosen STIS data reduction.
    %Given the dominant influence of NIRISS on the retrieved behavior of the non-gray cloud parametrization, tweaked reductions showing differences in trend were also tested. In all tested cases, the retrievals still favored a sub-Rayleigh slope.
    
    \item \textbf{Comparison to Information Content of WASP-39b} \\
    The obtained results were compared to previous information content studies of the benchmark hot Jupiter WASP-39b. Many results were found to be consistent such as the necessity of NIRISS data to obtain a preference for the non-gray cloud parametrization, and the insufficiency of NIRSpec alone for a significant \ce{H2S} detection. There are, however, notable differences such as the lack of a preference for a non-isothermal $T$--$P$ profile for any of the HAT-P-12b retrievals, and the minimal impact of the assumed $T$--$P$ structure on molecular abundances. These likely reflect genuine differences in the atmospheric temperature structures of the two planets.
\end{enumerate}

Several aspects remain unexplored in both the data reduction and retrieval modeling. These include accounting for potential limb asymmetries in the transit lightcurves, which could provide constraints on atmospheric variations along the terminator, as well as fitting for unocculted stellar heterogeneities during the retrievals. Furthermore, all molecular abundances reported in this study were derived %assuming \enquote{free chemistry}, i.e. 
without imposing any constraints on the underlying chemical and physical processes. A complementary analysis assuming equilibrium chemistry and additional, various disequilibrium effects would help assess the robustness of the obtained results. Extending such studies to additional exoplanets beyond WASP-39b and HAT-P-12b, particularly from different regions of the exoplanet parameter space, is essential for building a more general understanding of the information content of JWST transmission spectroscopy and obtaining robust constraints on their atmospheric properties.
%Lastly, acquiring further observations in the NIRSpec G395M wavelength range, such as with NIRSpec PRISM, could help resolve remaining discrepancies between data reductions.
The analysis of the JWST data of HAT-P-12b will be continued in Bouwman et~al. (in prep.), which will present the MIRI LRS data reduction, discuss the spectral features observed (or absent) in that wavelength range in comparison to similar targets such as WASP-107b \citep{Dyrek_2024_SO2SilicateClouds}, and explore different cloud treatments.

%\newpage

\begin{acknowledgements}
\input{acknowledgements}
\end{acknowledgements}

\bibliography{references}

%\newpage
\begin{appendix}

\onecolumn
\section{Standard reduction}\label{app_sec:data_reduct}

\begin{table*}[h]\label{tab:reduction_params}
    \caption{Parameters used for the standard reduction and retrievals, with priors and posteriors from the white lightcurve fit.}
    %\resizebox{1.0\textwidth}{!}{}
    \centering
    \input{tables/reduction_params}
    \tablefoot{
    Uncertainties are given in parentheses for brevity and correspond to the $1\sigma$ errors in the last digits.
    \textit{Priors:} $\mathcal{U}$ = Uniform, $\mathcal{N}$ = Normal, $\mathcal{L}$ = Log-Uniform; \textit{References:} [1] \citet{Mancini_2018_GAPSProgrammeHARPSN}, [2] \citet{Kokori_2022_ExoClockProjectII}; \textit{Spectral lightcurves:} [A] Same prior used, [B] Fixed to median value from white lightcurves, [C] Fixed to values from theoretical models; \textit{Minor Remarks:} (a) Used for wavelength calibration \& calculation of spectral lightcurve LDCs, (b) Value of 4.5 used for spectral LDCs due to grid limitations, (c) $\mathrm{BMJD} = \mathrm{BJD} - 2400000.5$, (d) Wide uninformative prior, (e) Prior calculated from literature values, (f) Assumption of circular orbit \citep[based on][]{Bonomo_2017_GAPSProgrammeHARPSN}, (g) Use of \citet{Kipping_2013_EfficientUninformativeSampling} parametrization.}
\end{table*}

\begin{figure*}[h!]
    \centering
    \includegraphics[width=\textwidth]{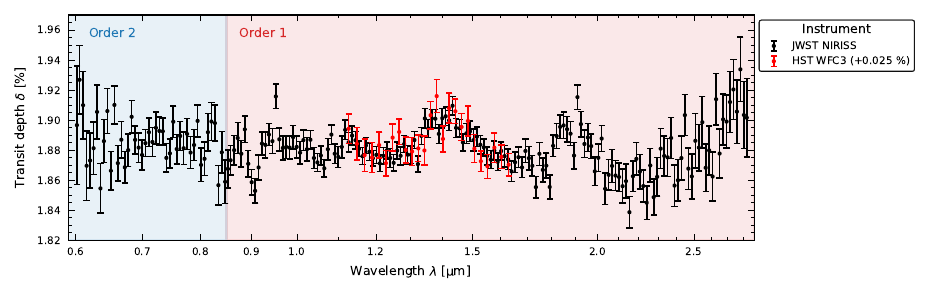}
    \caption{Comparison between the JWST NIRISS transmission spectrum from the standard reduction and the 
    HST WFC3 spectrum from the CASCADe reduction \citep{Crouzet_2025_DetectionCO$_2$CO}. A small offset of \SI{0.025}{\percent} was added to the HST WFC3 spectrum to allow for a better comparison of the spectral shapes.}\label{fig:transspec_comparison}
\end{figure*}

\FloatBarrier
\twocolumn

\begin{figure}
    \centering
    \includegraphics[width=\columnwidth]{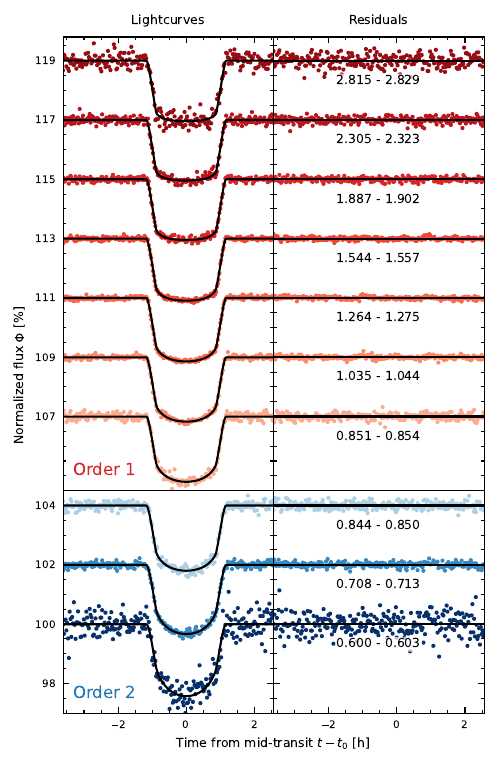}
    \caption{Representative spectral lightcurves of the NIRISS SOSS data with best-fit models (left) and residuals (right), spanning both spectral orders. Wavelength ranges (\si{\micro\meter}) are annotated, and integer-percent offsets were applied to separate the bins.}\label{fig:spectral_lcurves}
\end{figure}

\section{Reduction tweaks}\label{app_sec:reduction_tweaks}

\begin{figure}
    \centering
    \includegraphics[width=\columnwidth]{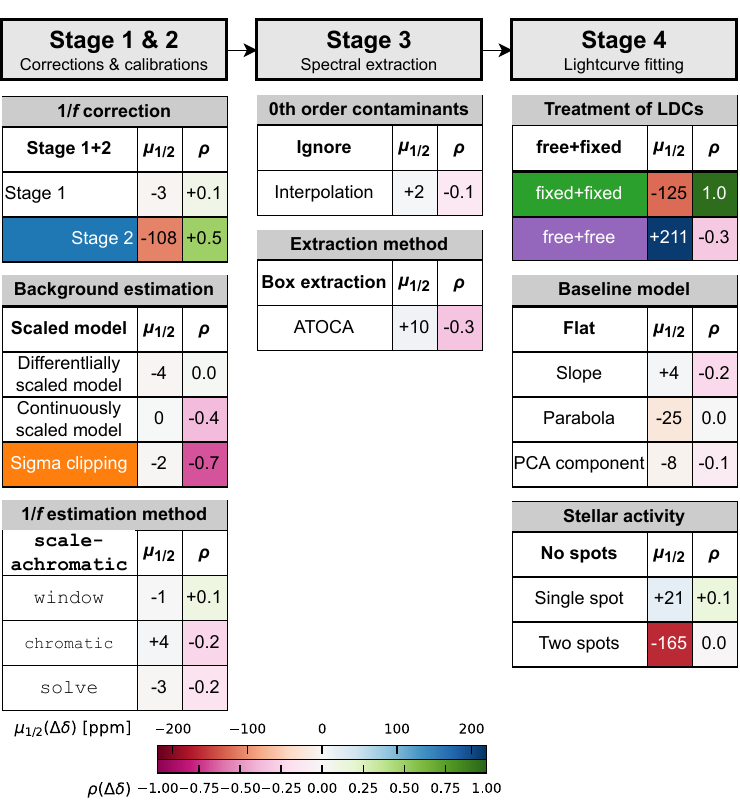}
    \caption{Schematic overview of the impact of the tested tweaked data reductions on the resulting transmission spectrum. Two metrics, computed from the transit-depth differences relative to the standard reduction ($\Delta\delta$), are reported: the weighted median $\mu_{1/2}$ and the Spearman rank correlation coefficient $\rho$, both listed and color-coded following the accompanying color bar. Tweaks are grouped by pipeline stage and ordered by execution, with the chosen method given on the left; the standard reduction choice is highlighted in bold, and select tweaked transmission spectra with large differences in trend are color-coded to match Fig.~\ref{fig:tweaks_impact}).}\label{fig:reduction_tweaks}
\end{figure}

Figure~\ref{fig:reduction_tweaks} provides a schematic overview of all the applied tweaks which will be described in the following section. It also lists quantities derived from the differences in the resulting transmission spectra, that were calculated to quantify the impact on the spectrum. They will be discussed in detail in the second subsection.

\subsection{Description of tweaks}\label{app_ssec:tweaks_description}

The only tweaks that were applied in the first two stages of the lightcurve extraction were related to the combined sky background and $1/f$ noise removal.
First, the effect of applying the $1/f$ noise correction to only the first or the second stage was tested.
Second, three alternative methods for estimating the sky background were explored.
The first two of these used alternative methods for determining scaling factors for the model background from the median stack of the out-of-transit images.
One method, which is directly implemented in ExoTEDRF, applies a differential scaling by introducing a second region to account for areas of the detector affected by dispersed zodiacal light. The default region was used ($\mathrm{x}_2 \in [715, 749]$, $\mathrm{y}_2 \in [235, 249]$; see Fig.\ref{fig:backs+contams}).
The other method applies a continuous scaling to each detector column. The custom scaling factors were derived by averaging the flux in a broader region of low target flux located above the third spectral order, followed by fitting a second-order polynomial to the column-averaged values. Specifically, all pixels located 40 pixels above the spectral trace were used, restricting the fit to columns with $\mathrm{x} < 1000 \, \mathrm{px}$. Columns with $\mathrm{x} < 230 \, \mathrm{px}$ were explicitly excluded due to contamination from the fourth spectral order \citep{Baines_2024_JWSTNIRISSSOSS}.
In addition, a region with a width of 30 pixels around the step feature was also excluded from the fit.
The third method derived a background estimate based solely on the median stack of the HAT-P-12b NIRISS observations. This was done by applying iterative sigma clipping ($\sigma=4$) to identify the pixels least affected by target flux, which were subsequently again column-averaged. Regions identified by the procedure as non-background have been highlighted in Figure~\ref{fig:backs+contams}. This method mirrors the approach used in the CASCADe reductions of the NIRSpec and MIRI data (\citealt{Crouzet_2025_DetectionCO$_2$CO}; Bouwman et~al., in prep.).
Third, the remaining three $1/f$ noise estimation methods implemented in ExoTEDRF were tested.
The first, \texttt{scale-achromatic-window}, modifies the standard reduction method by 
restricting the pixels considered for the $1/f$ noise estimation to those within a specific radial range around the spectral traces.
The second method \texttt{scale-chromatic} modifies the previous approach by using the first pass estimate of the spectral instead of the white lightcurves. 
The second, \texttt{scale-chromatic}, follows the same approach but uses the first pass spectral lightcurve instead of the white lightcurve.
The third, \texttt{solve}, determines the scaling of the median stack without prior assumptions by simultaneously solving for both the scaling factor and the $1/f$ noise amplitude.
Further details are provided in the pipeline documentation\footnote{\url{https://exotedrf.readthedocs.io/en/latest/content/usage.html\#a-note-on-1-f-correction-methods}}.

As noted earlier, no F277W exposure was obtained, preventing a direct measurement of contaminants isolated from the target flux.
To still assess the impact of potential background sources, the data were visually inspected for signatures of contamination. No dispersed light from non-target sources was identified.
Several undispersed \enquote{0th order} background sources are present, and two potential such contaminants were found to overlap with the cores of the target's first and second spectral orders (see Fig.~\ref{fig:backs+contams}).
A custom interpolation method was employed to reconstruct the affected regions, using Gaussian-weighted averages of neighboring pixels on either side of each contaminant over a range of $\Delta x = 20,\mathrm{px}$ along the dispersion axis and within $r_\mathrm{y} \pm 20,\mathrm{px}$ around the trace centroid along the spatial axis.
The interpolated values were linearly blended across the contaminated columns. An additional tweak applied in the first part of the reduction involved the use of the alternative ATOCA extraction algorithm, which models cross-contamination between spectral orders \citep{Darveau-Bernier_2022_ATOCAAlgorithmTreat}.

\begin{figure*}[h!]
    \centering
    \includegraphics[width=\textwidth]{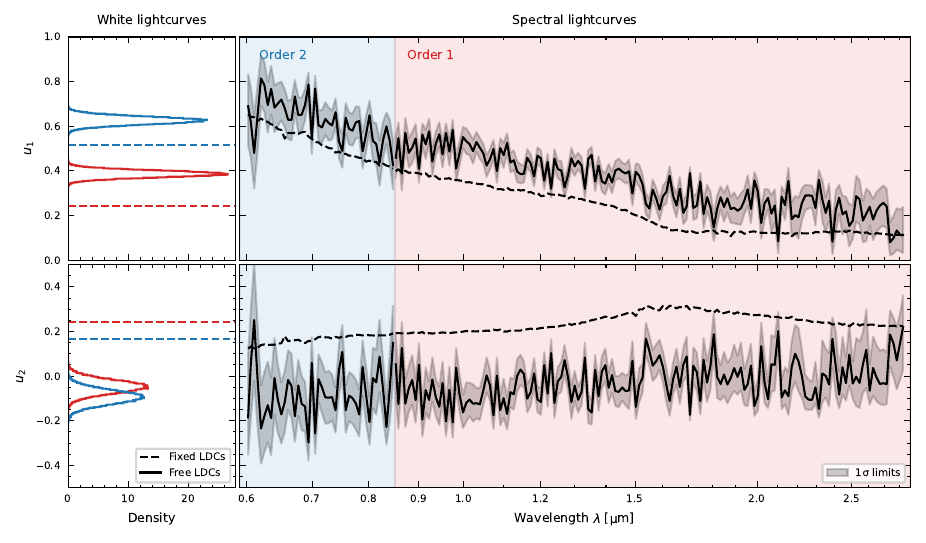}
    \caption{Comparison of the quadratic limb darkening coefficients (LDCs) $u_1$ and $u_2$ obtained when fitting them as free parameters (dashed lines) or when fixing them to values calculated with the \texttt{ExoTiC-LD} package \citep{Grant_2024_ExoTiCLDThirtySeconds} using the Stagger model grid \citep{Magic_2015_StaggergridGrid3D} (solid lines), shown for both the white and spectral lightcurves.
    Color is used to distinguish between the first (red) and second (blue) spectral orders. For the white lightcurves, the full marginal posterior distributions of the LDCs are shown, while for the spectral lightcurves the median values (lines) and $1\sigma$ limits (gray shading) are displayed.
    In cases with free LDCs, the \citet{Kipping_2013_EfficientUninformativeSampling} parametrization ($q_1$, $q_2$) was used, with samples transformed to $u_1$ and $u_2$ for visualization. The free LDCs for the spectral lightcurves were obtained using the orbital parameters from the white lightcurve fit, in which the LDCs were also fitted freely (free+free).}\label{fig:limb_darkening}
\end{figure*}

For the lightcurve fitting, a number of tweaks affecting the lightcurve shape, but not directly originating from the planet, were tested.
First, the impact of the adopted treatment of the LDCs was assessed by either fixing them to model values already at the white lightcurve level (fixed+fixed) or allowing them to vary freely within the \citet{Kipping_2013_EfficientUninformativeSampling} parametrization also at the spectral lightcurve level (free+free).
The differences between fitted and model-based LDC values are shown in Fig.~\ref{fig:limb_darkening} for both the white and spectral lightcurves. In both cases, clear and systematic offsets are present. For the white lightcurves, the fitted values favor higher $u_1$ and lower $u_2$ compared to the model values for both spectral orders: order~1 ($\Delta u_1 = +0.14 = +9.36\sigma$, $\Delta u_2 = -0.29 = -9.51\sigma$) and order~2 ($\Delta u_1 = +0.11 = +6.21\sigma$, $\Delta u_2 = -0.26 = -8.13\sigma$). A similar behavior is observed for the spectral lightcurves ($\Delta u_1 = 0.12_{-0.06}^{+0.05} = +1.9_{-1.1}^{+1.8}\sigma$, $\Delta u_2 = -0.27_{-0.08}^{+0.09} = -2.9_{-2.0}^{+1.5}\sigma$), where the relative trends between fitted and model values closely resemble those seen in the white lightcurves, differing primarily by an overall offset.
Second, alternative non-flat baseline models at varying levels of complexity were tested, namely a simple slope, a parabola and the use of the dominant ExoTEDRF-determined PCA component as a regressor. 
The parameters of the baseline models were allowed to vary freely and independently in their respective spectral lightcurve fits.
Third, the impact of stellar contamination from spot crossing events was assessed by coupling the \texttt{spotrod} \citep{Beky_2014_SPOTRODSemianalyticModel} code with the \texttt{juliet} framework, following an approach similar to the one presented in \citet{Fournier-Tondreau_2024_NearInfraredTransmissionSpectroscopy, Fournier-Tondreau_2025_TransmissionSpectroscopyWASP52} including the construction of a single combined white lightcurve from 
both spectral orders. Scenarios with one and two spot crossing events were tested. For the spectral lightcurves, all parameters, except for the spot contrast $f$, were fixed to the values obtained from the white lightcurve fit.

The configuration files, scripts, and notebooks used to perform the tweaked reductions, as well as the resulting transmission spectra, are publicly available (see Acknowledgements).

\subsection{Impact on transmission spectrum}\label{app_ssec:tweaks_impact}

Given the number of tested tweaks and to allow for a clear and efficient comparison, the differences in the transmission spectrum caused by any particular tweak were summarized using two statistical metrics. First, the differences between the transit depths at all wavelengths $\Delta\delta$ were calculated. From these differences, a weighted median $\mu_{1/2}$ using the inverse of the combined uncertainties as weights was calculated to quantify any general offset. Additionally, Spearman’s rank correlation coefficient $\rho$ was computed to assess differences in the overall trend. The resulting values are listed in the diagram in Figure~\ref{fig:reduction_tweaks}.

\begin{figure*}[h]
    \centering
    \includegraphics[width=0.9\textwidth]{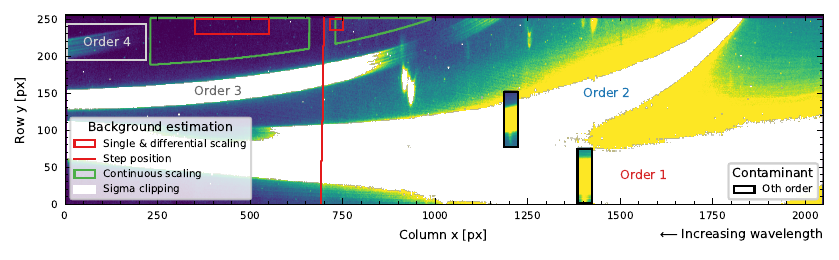}
    \caption{Overview of the data recorded by the NIRISS detector, based on a stack of out-of-transit integration images. The spectral orders are labeled, with a different color scale applied to enhance the visibility of the faint 4th order (gray rectangle). 
    Regions used for scaling the model background and the location of the background step are highlighted (red and green).
    Areas identified via sigma clipping as containing signal (rather than background) have been masked (white).
    The locations of two potential 0th-order contaminants overlapping with the spectral traces are also indicated. For better visibility, these regions are shown with a separate color scale and an enlarged box ($2 \times \Delta \mathrm{x}$) compared to the area that was interpolated over.}\label{fig:backs+contams}
\end{figure*}

\begin{figure*}[h!]
    \centering
    \includegraphics[width=\textwidth]{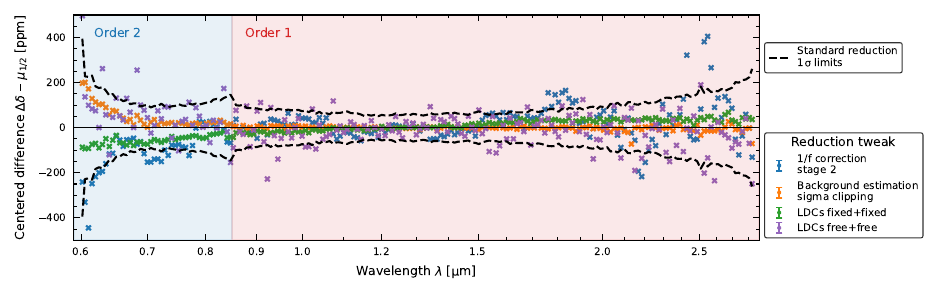}
    \caption{Impact of select reduction tweaks, shown as differences between the tweaked and standard-reduction transmission spectra. The differences were calculated by subtracting the weighted median $\mu_{1/2}$ from the transit depth differences $\Delta\delta$. This centering removes global offsets and emphasizes wavelength-dependent trends. For comparison, the $1\sigma$ uncertainty range of the standard reduction is indicated via dashed lines.}\label{fig:tweaks_impact}
\end{figure*}

Only a few tweaks lead to offsets $\mu_{1/2} > 100 \, \mathrm{ppm}$ (approx. mean uncertainty of standard reduction), or yield a Spearman coefficient $|\rho| \gtrsim 0.2$, indicating a strong wavelength-dependent trend. Four exemplary reductions with particularly large correlation coefficients (highlighted in Fig.~\ref{fig:reduction_tweaks}) are shown in Figure~\ref{fig:tweaks_impact}. 
Given that the retrievals freely fit for an offset between the different instruments, a pure shift in the spectral baseline is not expected to influence the retrieval results. Therefore, the differences were centered by subtracting the previously calculated weighted median values.
There is a general trend of increased discrepancies ($\gtrsim 200 \, \mathrm{ppm}$) at both of the noisier ends of the spectrum, with the shortest wavelengths of the second order being affected more strongly. Except for two tweaked reductions that exhibit generally higher localized scatter, the centered differences remain within the $1\sigma$ uncertainty limits of the standard reduction.
Applying the $1/f$ noise correction only in the second stage results not only in a large overall offset ($\sim 100 \, \mathrm{ppm}$), but also in localized differences across the entire wavelength range, with particularly pronounced deviations at its extremes (up to $\sim 400 \, \mathrm{ppm}$). In contrast, no such large discrepancies were observed when the $1/f$ noise correction was only applied in the first stage. This highlights the importance of performing the $1/f$ noise correction already at the group level, rather than postponing this correction to the integration level where the noise from multiple groups has been combined.
For the background estimation, the sigma clipping method most strongly affects the overall shape of the transmission spectrum; notable differences of $\rho =-0.4$ also arise when using the custom scaling of the model background. Both result in a steeper slope at the short-wavelength end. This region is particularly sensitive due to two complementary factors: Firstly, the second spectral order reaches its short-wavelength limit in the strongly illuminated right part of the detector (cf. Fig.~\ref{fig:backs+contams}). Secondly, the flux of the second order is decreasing towards shorter wavelengths leading to an increasing dominance of the flux from the first order despite increasing trace separation (maximum in flux ratio and ATOCA estimate of cross-contamination).
Neither tweak introduces a significant global offset.
The alternative treatments of the LDCs by keeping them free or fixed for the white and spectral lightcurves both have significant impacts on offset ($|\mu_{1/2}| \geq 125\, \mathrm{ppm}$) and trend ($|\rho| \geq 0.3$). This is consistent with the well-established finding that strong degeneracies between LDCs and orbital parameters, such as the impact parameter $b$, can introduce significant biases \citep{Espinoza_2015_LimbDarkeningExoplanets, Alexoudi_2020_RoleImpactParameter}. These can also simply arise from the chosen parametrization itself and can cause significant wavelength-dependent biases across JWST modes, including NIRISS SOSS \citep{Coulombe_2024_BiasesExoplanetTransmission}.

The choice of the baseline model does not introduce significant trends or offsets. Fitting for a linear slope ($\Delta\ln Z = -2.05 \pm 0.18$), a quadratic parabola ($\Delta\ln Z = -4.36 \pm 0.18$), or the dominant PCA component determined by the ExoTEDRF pipeline ($\Delta\ln Z = -11.97 \pm 0.17$) all result in lower Bayesian evidence, indicating a clear preference for the simpler flat baseline model.
In contrast, spot modeling reveals a strong preference for the inclusion of a single spot ($\Delta\ln Z = +8.85 \pm 0.14$), while modeling two spots is disfavored ($\Delta\ln Z = -1.73 \pm 0.14$).
While there is a larger offset in the two spot case ($|\mu_{1/2}| = 165 \, \mathrm{ppm}$) it remains very moderate in the single spot case ($|\mu_{1/2}| = 21 \, \mathrm{ppm}$).
The difference in the shape of the resulting transmission spectra, however, remains small in both cases ($|\rho| \leq 0.1$).
Retrievals using all JWST data but replacing the NIRISS spectrum with that from the tweaked reductions yield negligible changes for most detected molecules ($< 0.07$~dex, $\lesssim 0.11 \sigma$) except for \ce{H2S} ($\lesssim 0.2$~dex, $\lesssim 0.5 \sigma$), and only slightly larger yet still minor differences ($\lesssim 0.3$~dex, $< 0.08 \sigma$) for non-detected molecules.

\onecolumn

\section{Retrievals}

\begin{figure*}[h]
    \centering
    \includegraphics[width=\textwidth]{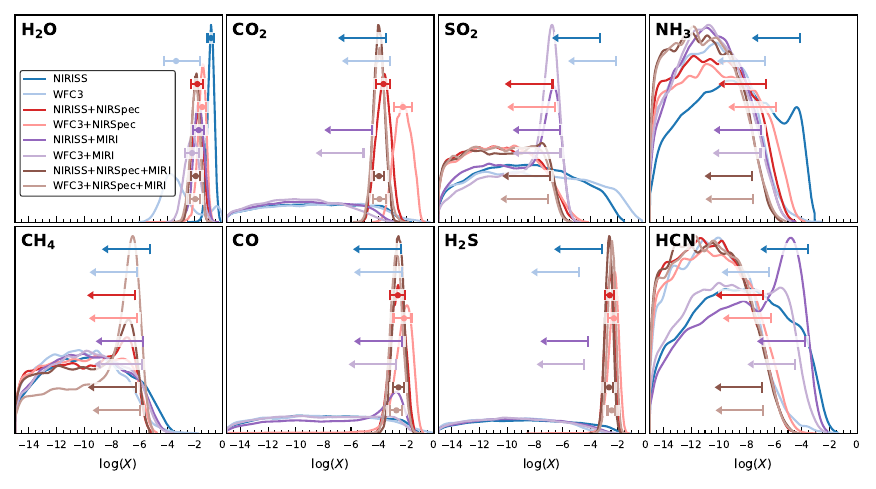}
    \caption{Same as Fig.~\ref{fig:marg_abunds_jwst}, but instead of all JWST combinations only the ones that include NIRISS SOSS are shown and compared to the results obtained when HST WFC3 is used instead.}\label{fig:marg_abunds_wfc3}
\end{figure*}

\begin{table*}[h]\label{tab:marg_abunds_jwst}
    \caption{Retrieved molecular abundances for all possible combinations of JWST instruments.}
    %\resizebox{1.0\textwidth}{!}{}
    \centering
    \input{tables/marg_abunds_jwst}
    \tablefoot{Median values with asymmetric $1\sigma$ uncertainties are given for cases where a constrained marginal posterior abundance was obtained; otherwise, the upper 95th-percentile limit is reported.}
\end{table*}

\begin{figure*}[h!]
    \centering
    \includegraphics[width=\textwidth]{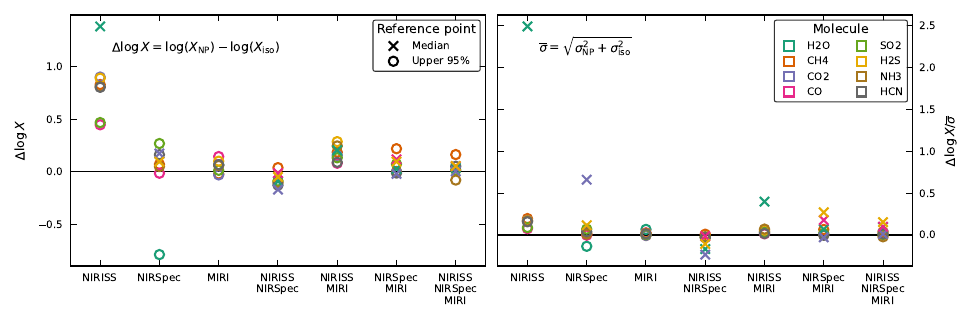}
    \caption{Differences in the retrieved chemical abundances between the standard $N$-point profile ($X_\mathrm{NP}$) and a simpler isothermal profile ($X_\mathrm{iso}$).
    In the left panel, the differences are shown as absolute differences in the logarithmic VMRs $\Delta\log X$, whereas in the right panel, the differences have been scaled by the combined uncertainty $\bar{\sigma}$ calculated by adding the individual $1\sigma$ interval widths of the VMRs in quadrature.
    For the cases of significant detections the median abundance was used for the comparison, while for non-detections the upper 95th percentile served as the reference point.}\label{fig:comp_abund_isothermal}
\end{figure*}

\FloatBarrier

\subsection{Impact of offset treatment}\label{app_ssec:offset_treatment}

\begin{figure*}[h!]
    \centering
    \includegraphics[width=\textwidth]{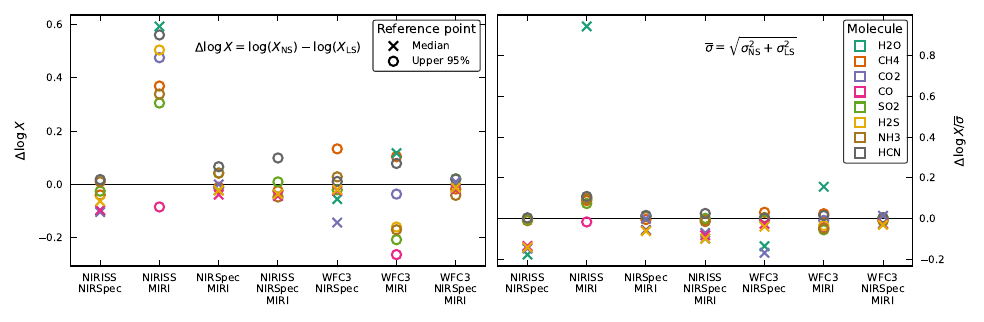}
    \caption{
    Same as Fig.~\ref{fig:comp_abund_isothermal}, showing the abundance differences between retrievals with instrumental offsets included as free parameters in the nested sampling ($X_\mathrm{NS}$) and those optimized separately via least-squares ($X_\mathrm{LS}$).
    All instrument combinations considered in this work, except those including HST STIS data, are shown.
    }\label{fig:comp_abund_offset}
\end{figure*}

\twocolumn

\noindent
While all multi-instrument retrievals conducted in this study include instrumental offsets between datasets, two different treatments were adopted. For all retrievals except those involving HST STIS data, the offsets $\Delta\delta_\mathrm{0,Z}$ were included as free parameters in the nested sampling. For retrievals involving STIS data, the offsets were instead determined via a separate least-squares optimization outside the nested sampling, following the approach also used for other ARCiS retrievals in earlier work in this series \citep{Crouzet_2025_DetectionCO$_2$CO}.
Including the offsets directly in the nested sampling is, in principle, the more consistent approach, as it accounts for potential degeneracies with atmospheric parameters such as molecular abundances and cloud opacity parameters. Additionally, optimizing the offsets outside the nested sampling can bias the Bayesian evidences, and thus the inferred detection significances, since it introduces parameters that are not explored within the sampling. However, this treatment comes with a significant computational overhead. While the increase in the number of model evaluations is typically modest (up to a factor of \num{\sim 2}) for two-instrument combinations, it can reach a factor of \num{\sim 4} for three-instrument combinations (e.g. NIRISS+NIRSpec+MIRI: from \num{\sim 4.2e6} to \num{\sim 1.8e7} model evaluations).
This substantial computational overhead is the reason why the alternative least-squares offset treatment was adopted for the retrievals including the HST STIS data.
These retrievals involve the largest number of independent datasets (four in total) and are therefore expected to incur the largest increase in total runtime due to the chosen offset treatment.
Combined with the increased per-model evaluation time due to the larger number of datapoints, this renders these retrievals computationally unfeasible within the scope of this work.

To compensate for this limitation, retrievals using the least-squares optimization were also performed for all other instrument combinations, enabling an assessment of the impact of the offset treatment on the retrieved parameters and Bayesian evidences.
Differences in the retrieved parameters and detection significances remain generally minor, with the only notable exceptions being the NIRISS+MIRI combination, and to a much lesser extent WFC3+MIRI. These are the only instrument combinations with a significant wavelength gap (\SI{>1.9}{\micro\meter}), compared to the otherwise near-continuous coverage (\SI{<0.03}{\micro\meter}), with the exception of the gap between WFC3 and NIRSpec at \SI{\sim1.2}{\micro\meter}.
The least-squares treatment yields spurious tentative evidence for CO in the NIRISS+MIRI combination ($2.6\sigma$), and a weak preference of similar significance for the $N$-point over the isothermal $T$--$P$ profile.
As shown in Fig.~\ref{fig:comp_abund_offset}, this is also the only combination for which differences in the marginal chemical abundances between the two methods exceed \num{0.3}~dex, with WFC3+MIRI showing the next-largest deviations, exceeding those of all remaining combinations by more than \num{0.1}~dex. Scaled by the uncertainties of the marginal abundances, however, the differences remain small, below $0.2\sigma$ in all cases, with the sole notable outlier being \ce{H2O} for the NIRISS+MIRI combination at $\sim 0.9\sigma$.
While this demonstrates that the choice of offset treatment can introduce systematic differences, supporting the inclusion of instrumental offsets in the nested sampling, the results also show that these differences remain minor in the absence of large wavelength gaps. Given the lack of such gaps in the retrievals involving the STIS data and the negligible impact of the low-resolution optical HST STIS data on molecular abundances, both in practice and as expected from molecular opacities, this justifies the adopted trade-off between computational cost and statistical rigor.

\end{appendix}
\end{document}

%% file: authors.tex
\author{
L. Heinke            \inst{1,2,3} \and % Linus Heinke      0000-0002-6826-9394
M. Min               \inst{4}     \and % Michiel Min       0000-0001-5778-0376
J. Bouwman           \inst{5}     \and % Jeroen Bouwman    0000-0003-4757-2500
N. Crouzet           \inst{6,7}   \and % Nicolas Crouzet   0000-0001-7866-8738
T. Konings           \inst{1}     \and % Thomas Konings    0000-0001-5849-9762
L. Decin             \inst{1}     \and % Leen Decin        0000-0002-5342-8612
L. B. F. M. Waters   \inst{8,4}   \and % Laurentius Bernardus Franciscus Maria (Rens) Waters 0000-0002-5462-9387
P.-O. Lagage         \inst{9}     \and % Pierre-Olivier Lagage
T. Henning           \inst{5}     \and % Thomas Henning    0000-0002-1493-300X
P. I. Palmer         \inst{2,3}   \and % Paul I. Palmer    0000-0002-1487-0969
B. Edwards           \inst{4}     \and % Billy Edwards     0000-0002-5494-3237
J. P. Pye            \inst{10}    \and % John P. Pye       0000-0002-0932-4330
M. Güdel             \inst{11,12} \and % Manuel Guedel     0000-0001-9818-0588 --> Use of Güdel instead of Guedel due to example in Schleich+2024
% ----------------------------------------------------------------------------------------- %
O. Absil             \inst{13}    \and % Olivier Absil     0000-0002-4006-6237
D. Barrado           \inst{14}    \and % David Barrado Navascués 0000-0002-5971-9242 --> Dropped last name like in Nature publication on 15NH3
C. Cossou            \inst{15}    \and % Christophe Cossou 0000-0001-5350-4796
A. Glasse            \inst{16}    \and % Alistair Glasse   0000-0002-2041-2462
A. M. Glauser        \inst{12}    \and % Adrian M. Glauser 0000-0001-9250-1547
G. Östlin            \inst{17}    \and % Göran Östlin      0000-0002-3005-1349
N. Whiteford         \inst{18}    \and % Niall Whiteford   0000-0001-8818-1544
T. P. Ray            \inst{19}         % Tom P. Ray        0000-0002-2110-1068
}

%% file: affiliations.tex
\institute{
% 1
Institute of Astronomy, KU Leuven, Celestijnenlaan 200D, 3001 Leuven, Belgium \and
% 2
School of GeoSciences, University of Edinburgh, Edinburgh, EH9 3FF, UK \and
% 3
Centre for Exoplanet Science, University of Edinburgh, Edinburgh, EH9 3FD, UK \and
% 4
SRON Space Research Organisation Netherlands, Niels Bohrweg 4, 2333 CA Leiden, The Netherlands \and
% 5
Max-Planck-Institut für Astronomie (MPIA), Königstuhl 17, 69117 Heidelberg, Germany \and
% 6
Leiden Observatory, Leiden University, P.O. Box 9513, 2300 RA Leiden, The Netherlands \and
% 7
Kapteyn Astronomical Institute, University of Groningen, P.O. Box 800, 9700 AV Groningen, The Netherlands \and
% 8
Department of Astrophysics/IMAPP, Radboud University, P.O. Box 9010, 6500 GL Nijmegen, The Netherlands \and
% 9
Université Paris-Saclay, Université Paris Cité, CEA, CNRS, AIM, F-91191 Gif-sur-Yvette, France \and
% 10
School of Physics \& Astronomy, Space Park Leicester, University of Leicester, 92 Corporation Road, Leicester, LE4 5SP, UK \and
% 11
Department of Astrophysics, University of Vienna, Türkenschanzstr. 17, 1180 Vienna, Austria \and
% 12
ETH Zürich, Institute for Particle Physics and Astrophysics, Wolfgang-Pauli-Str. 27, 8093 Zürich, Switzerland \and
% 13
STAR Institute, Université de Liège, Allée du Six Août 19c, 4000 Liège, Belgium \and
% 14
Centro de Astrobiología (CAB), CSIC-INTA, ESAC Campus, Camino Bajo del Castillo s/n, 28692 Villanueva de la Cañada, Madrid, Spain \and
% 15
Université Paris-Saclay, CEA, IRFU, 91191, Gif-sur-Yvette, France  \and
% 16
UK Astronomy Technology Centre, Royal Observatory, Blackford Hill, Edinburgh EH9 3HJ, UK  \and
% 17
Department of Astronomy, Oskar Klein Centre, Stockholm University, 106 91 Stockholm, Sweden  \and
% 18
Department of Astrophysics, American Museum of Natural History, New York, NY 10024, USA \and
% 19
School of Cosmic Physics, Dublin Institute for Advanced Studies, 31 Fitzwilliam Place, Dublin, D02 XF86, Ireland
}

%% file: tables/observations.tex
\begin{tabular}{clcccccllc}
\hline\hline
Telescope             & Instrument & Mode & Grating                                                       & \begin{tabular}[c]{@{}c@{}}Wavelength \\ $\lambda$ [$\si{\micro\meter}$]\end{tabular} & \begin{tabular}[c]{@{}c@{}}Resolution\\ $\lambda/\Delta\lambda$\end{tabular} & \begin{tabular}[c]{@{}c@{}}Uncertainty\\ $\sigma_{\delta}$ [ppm]\end{tabular} & Obs. date                                                                    & PID                                                                                                   & Ref.   \\ \hline
\multirow{3}{*}{JWST} & NIRISS     & SOSS & --                                                            & $0.60 \, \text{--} \, 2.83$                                                           & $125.0 \pm 1.4$                                                              & $100_{-50}^{+120}$                                                            & 2023-06-20                                                                   & \multirow{3}{*}{\href{https://www.stsci.edu/jwst-program-info/program/?program=1281&submit=Go}{1281}} & [0]    \\
                      & NIRSpec    & BOTS & G395M                                                         & $2.84 \, \text{--} \, 5.18$                                                           & $143 \, \text{--} \, 259$                                                    & $140_{-60}^{+110}$                                                            & 2023-02-11                                                                   &                                                                                                       & [1]    \\
                      & MIRI       & LRS  & --                                                            & $4.6 \, \text{--} \, 12.4$                                                            & $24 \, \text{--} \, 62$                                                      & $290_{-170}^{+740}$                                                           & 2023-06-07                                                                   &                                                                                                       & [2]    \\ \hline
\multirow{2}{*}{HST}  & STIS       &      & \begin{tabular}[c]{@{}c@{}}G430L\\ G430L\\ G750L\end{tabular} & $0.36 \, \text{--} \, 1.01$                                                           & $8 \, \text{--} \, 36$                                                       & $310_{-130}^{+424}$                                                           & \begin{tabular}[c]{@{}l@{}}2012-04-11\\ 2012-04-30\\ 2023-02-04\end{tabular} & \href{https://www.stsci.edu/hst-program-info/program/?program=12473&submit=Go}{12473}                 & [3][4] \\
                      & WFC3       & IR   & G141                                                          & $1.12 \, \text{--} \, 1.64$                                                           & $52 \, \text{--} \, 71$                                                      & $91_{-19}^{+18}$                                                              & \begin{tabular}[c]{@{}l@{}}2015-12-12\\ 2016-08-31\end{tabular}              & \href{https://www.stsci.edu/hst-program-info/program/?program=14260&submit=Go}{14260}                 & [1]    \\ \hline
\end{tabular}

%% file: tables/retrieval_priors.tex
\begin{tabular}{lccc}
\hline\hline
Name                 & Symbol                      & Unit                & Prior                           \\ \hline
Planetary radius     & $R_\mathrm{p}$              & $\mathrm{R_{Jup}}$  & $\mathcal{U}(0.5, 1.5)$         \\
Temperature point    & $T_\mathrm{p,i}$            & $\si{\kelvin}$      & $\mathcal{U}(100, 1500)$        \\
Chemical abundance   & $X_\mathrm{X}$              &                     & $\mathcal{L}(10^{-15}, 10^{0})$ \\
Instrumental offset  & $\Delta\delta_\mathrm{0,Z}$ & ppm                 & $\mathcal{U}(-1000, +1000)$     \\ \hline
\multicolumn{4}{c}{Cloud parametrization}                                                                  \\ \hline
Cloud-top pressure   & $P_\mathrm{cloud}$          & $\si{\bar}$         & $\mathcal{L}(10^{-10}, 10^{2})$ \\
Mixing ratio         & $X_\mathrm{cloud}$          &                     & $\mathcal{L}(10^{-15}, 10^{0})$ \\
Reference wavelength & $\lambda_{0}$               & $\si{\micro\meter}$ & $\mathcal{L}(10^{-2}, 10^{2})$  \\
Spectral slope       & $p$                         &                     & $\mathcal{U}(0, 6)$             \\ \hline
\end{tabular}

%% file: acknowledgements.tex
We thank M. Radica and M. Fournier-Tondreau for valuable discussions and advice regarding the NIRISS SOSS data reduction. We also thank B. Béky for technical assistance with the \texttt{spotrod} code.
%MPIA acknowledges support from the Federal Ministry of Economy (BMWi) through the German Space Agency (DLR) and the Max Planck Society.
L.H. and L.D. acknowledge funding from the European Union H2020-MSCA-ITN-2019 under Grant no. 860470 (CHAMELEON).
L.H., T.K., and L.D. acknowledge funding from the KU Leuven Interdisciplinary Grant (IDN/19/028).
L.H. acknowledges funding from the FWO research grant G0B3823N.
L.H. and L.D. acknowledge funding from the KU Leuven Methusalem Grant SOUL (METH/24/012).
L.D. acknowledges funding from the FWO research grant G086217N.
P.-O.L. and C.O. acknowledge funding support by CNES.
J.P.P. acknowledges financial support from the UK Science and Technology Facilities Council, and the UK Space Agency. For the purpose of open access, the author has applied a Creative Commons Attribution (CC BY) licence to the Author Accepted Manuscript version arising from this submission.
O.A. is a Senior Research Associate of the Fonds de la Recherche Scientifique – FNRS. O.A. thanks the European Space Agency (ESA) and the Belgian Federal Science Policy Office (BELSPO) for their support in the framework of the PRODEX Programme.
D.B. has been funded by grant No. PID2023-150468NB-I00 by the Spain Ministry of Science, Innovation/State Agency of Research MCIN/AEI/ 10.13039/501100011033.
Support from SNSA is acknowledged.
N.W. acknowledges support from NSF awards \#2238468 and \#1909776, and from NASA award \#80NSSC22K0142.
T.P.R. acknowledges support from the ERC 743029 EASY.
\\
\textit{Facilities:}
This work is based on observations made with the NASA/ESA/CSA James Webb Space Telescope. The data were obtained from the Mikulski Archive for Space Telescopes at the Space Telescope Science Institute, which is operated by the Association of Universities for Research in Astronomy, Inc., under NASA contract NAS 5-03127 for JWST. These observations are associated with program \#1281.
This research is based on observations made with the NASA/ESA \textit{Hubble Space Telescope} obtained from the Space Telescope Science Institute, which is operated by the Association of Universities for Research in Astronomy, Inc., under NASA contract NAS 5–26555. These observations are associated with programs 12473 \& 14260.
\\
\textit{Databases:}
This research has made use of the NASA Exoplanet Archive, which is operated by the California Institute of Technology, under contract with the National Aeronautics and Space Administration under the Exoplanet Exploration Program.
We acknowledge the use of the ExoAtmospheres database during the preparation of this work.
\\
\textit{Software:}
General analysis made use of 
\texttt{NumPy} \citep{Harris_2020_ArrayProgrammingNumPy}, 
\texttt{Matplotlib} \citep{Hunter_2007_Matplotlib2DGraphics}, 
\texttt{corner} \citep{Foreman-Mackey_2016_CornerpyScatterplotMatrices}, 
and \texttt{Astropy} \citep{AstropyCollaboration_2022_AstropyProjectSustaining}. 
Data reduction employed 
\texttt{ExoTEDRF} \citep{Radica_2024_ExoTEDRFEXOplanetTransit}, 
\texttt{jwst} \citep{bushouse_2023_10022973}, 
\texttt{juliet} \citep{Espinoza_2019_JulietVersatileModelling}, 
\texttt{ExoTiC-LD} \citep{Grant_2024_ExoTiCLDThirtySeconds}, 
\texttt{batman} \citep{Kreidberg_2015_BatmanBAsicTransit}, 
\texttt{spotrod} \citep{Beky_2014_SPOTRODSemianalyticModel}, 
and \texttt{dynesty} \citep{Speagle_2020_DYNESTYDynamicNested}. 
Atmospheric retrievals were performed with 
\texttt{ARCiS} \citep{Min_2020_ARCiSFrameworkExoplanet} 
coupled to \texttt{MultiNest} \citep{Feroz_2009_MULTINESTEfficientRobust}.
\\
\textit{Data availability:}
The configuration files, scripts, and notebooks used for the standard and tweaked NIRISS data reductions, together with all resulting transmission spectra, are publicly available via Zenodo (DOI: \zenododoi).
The transmission spectrum from the standard reduction is additionally available through CDS.
The ARCiS input files and posterior samples corresponding to the retrievals performed in this work are archived in the same Zenodo record.
The sources of the additional reduced JWST and HST datasets used in the retrievals are listed in Table~\ref{tab:observations}. All raw JWST and HST observations analyzed in this work are publicly available from the Mikulski Archive for Space Telescopes (MAST).

%% file: tables/reduction_params.tex
\begin{tabular}{lcccccc}
\hline\hline
Name                            & Symbol                                  & Unit                          & Order & Prior                                           & Posterior                      & Note                           \\ \hline
\multicolumn{7}{c}{Star}                                                                                                                                                                                                              \\ \hline
Mass                            & $M_\star$                               & $\mathrm{M_\odot}$            &       & 0.691                                           &                                & \multirow{3}{*}{[1], (a)}      \\
Radius                          & $R_\star$                               & $\mathrm{R_\odot}$            &       & 0.679                                           &                                &                                \\
Effective temperature           & $T_\mathrm{eff}^{\star}$                & \si{\kelvin}                  &       & 4665                                            &                                &                                \\
Surface gravity                 & $\log g_\star$                          & $\log(\mathrm{g \, cm}^{-3})$ &       & 4.614 (4.5)                                     &                                & [1], (a), (b)                  \\
Metallicity                     & $[\mathrm{Fe/H}]$                       &                               &       & -0.20                                           &                                & [1], (a)                       \\
Distance                        & $d$                                     & pc                            &       & 142.751                                         &                                &                                \\ \hline
\multicolumn{7}{c}{Orbit}                                                                                                                                                                                                             \\ \hline
Period                          & $P$                                     & \si{\day}                     &       & 3.21305751                                      &                                & [2], [A]                       \\
Transit mid-time                & $t_0$                                   & BMJD                          &       & 60115 + $\mathcal{U}(0.3, 0.6)$                 & 60115.447938(12)               & [B], (c), (d)                  \\
Semi-major axis                 & $a$                                     & au                            &       & 0.03767                                         &                                & [1]                            \\
Scaled semi-major axis          & $a/R_\star$                             &                               &       & $\mathcal{N}(11.9, 0.3)$                        & 11.69(3)                       & \multirow{2}{*}{[1], [B], (e)} \\
Impact parameter                & $b$                                     &                               &       & $\mathcal{N}(0.19, 0.05)$                       & 0.258(11)                      &                                \\
Eccentricity                    & $e$                                     &                               &       & 0                                               &                                & \multirow{2}{*}{[1], [A], (f)} \\
Argument of periastron          & $\omega$                                & \si{\degree}                  &       & 90                                              &                                &                                \\ \hline
\multicolumn{7}{c}{Planet}                                                                                                                                                                                                            \\ \hline
Mass                            & $M_\mathrm{p}$                          & $\mathrm{M_{Jup}}$            &       & 0.201                                           &                                & [1]                            \\
\multirow{2}{*}{Radius ratio}   & \multirow{2}{*}{$R_\mathrm{p}/R_\star$} &                               & 1     & \multirow{2}{*}{$\mathcal{U}(0.01, 0.9)$}       & 0.13794(17)                    & \multirow{2}{*}{[A], (d)}      \\
                                &                                         &                               & 2     &                                                 & 0.13825(23)                    &                                \\ \hline
\multicolumn{7}{c}{Limb darkening}                                                                                                                                                                                                    \\ \hline
\multirow{4}{*}{Quadratic LDCs} & \multirow{2}{*}{$q_1$}                  &                               & 1     & \multirow{2}{*}{$q_1 \in [0,1]$}                & 0.115(11)                      & \multirow{4}{*}{[C], (d), (g)} \\
                                &                                         &                               & 2     &                                                 & 0.282(17)                      &                                \\
                                & \multirow{2}{*}{$q_2$}                  &                               & 1     & \multirow{2}{*}{$q_2 \in [0,1]$}                & 0.57(5)                        &                                \\
                                &                                         &                               & 2     &                                                 & 0.59(3)                        &                                \\ \hline
\multicolumn{7}{c}{Systematic}                                                                                                                                                                                                        \\ \hline
Dilution factor                 & $D$                                     &                               & 1, 2  & 1                                               &                                & \multirow{4}{*}{[A]}           \\
Relative out-of-transit flux    & $M$                                     &                               & 1, 2  & $\mathcal{N}(0.0, 0.1)$                         & $\left| M \right|$ < 10$^{-5}$ &                                \\
\multirow{2}{*}{Jitter term}    & \multirow{2}{*}{$\sigma_\omega$}        & \multirow{2}{*}{ppm}          & 1     & \multirow{2}{*}{$\mathcal{L}(10^{-1}, 10^{4})$} & 95(5)                          &                                \\
                                &                                         &                               & 2     &                                                 & 146(10)                        &                                \\ \hline
\end{tabular}

%% file: tables/marg_abunds_jwst.tex
\begin{tabular}{lccccccccc}
\hline\hline
Instruments                                                       & $\log(X_{\mathrm{H}_2\mathrm{O}})$ &  & $\log(X_{\mathrm{CH}_4})$ & $\log(X_{\mathrm{CO}_2})$ & $\log(X_{\mathrm{CO}})$ & $\log(X_{\mathrm{SO}_2})$ & $\log(X_{\mathrm{H}_2\mathrm{S}})$ & $\log(X_{\mathrm{NH}_3})$ & $\log(X_{\mathrm{HCN}})$ \\ \hline
NIRISS                                                            & $-0.83_{-0.25}^{+0.23}$            &  & $< -5.26$                 & $< -3.49$                 & $< -2.40$               & $< -3.31$                 & $< -3.16$                          & $< -4.16$                 & $< -3.54$                \\
NIRSpec                                                           & $< -2.60$                          &  & $< -6.28$                 & $0.92_{-0.15}^{+0.16}$    & $< -2.40$               & $< -6.68$                 & $-3.12_{-0.53}^{+0.43}$            & $< -5.10$                 & $< -6.01$                \\
MIRI                                                              & $-2.26_{-0.79}^{+0.69}$            &  & $< -5.99$                 & $< -5.23$                 & $< -1.70$               & $< -6.58$                 & $< -4.29$                          & $< -7.06$                 & $< -3.72$                \\ \hline
\begin{tabular}[c]{@{}l@{}}NIRISS+\\ NIRSpec\end{tabular}         & $-1.82_{-0.44}^{+0.38}$            &  & $< -6.33$                 & $-3.65_{-0.53}^{+0.50}$   & $-2.61_{-0.55}^{+0.51}$ & $< -6.73$                 & $-2.58_{-0.34}^{+0.32}$            & $< -6.57$                 & $< -6.78$                \\
\begin{tabular}[c]{@{}l@{}}NIRISS+\\ MIRI\end{tabular}            & $-1.71_{-0.39}^{+0.39}$            &  & $< -5.71$                 & $< -4.47$                 & $< -2.32$               & $< -6.15$                 & $< -4.16$                          & $< -6.92$                 & $< -3.79$                \\
\begin{tabular}[c]{@{}l@{}}NIRSpec+\\ MIRI\end{tabular}           & $-1.83_{-0.30}^{+0.27}$            &  & $< -5.68$                 & $-3.79_{-0.41}^{+0.42}$   & $-2.74_{-0.49}^{+0.48}$ & $< -7.13$                 & $-2.30_{-0.28}^{+0.26}$            & $< -7.37$                 & $< -6.46$                \\ \hline
\begin{tabular}[c]{@{}l@{}}NIRISS+\\ NIRSpec+\\ MIRI\end{tabular} & $-1.93_{-0.32}^{+0.30}$            &  & $< -6.27$                 & $-3.97_{-0.38}^{+0.39}$   & $-2.57_{-0.41}^{+0.38}$ & $< -6.91$                 & $-2.64_{-0.28}^{+0.27}$            & $< -7.58$                 & $< -6.85$                \\ \hline
\end{tabular}